\documentclass[]{article}

\usepackage{amsmath}
\usepackage{amsfonts}
\usepackage{graphicx}
\usepackage{xcolor}
\usepackage[margin=1in]{geometry}
\usepackage{float}
\usepackage{bm}
\usepackage{theorem}
\usepackage{breqn}
\usepackage{subcaption}
\usepackage{appendix}
\usepackage[colorlinks,citecolor=red,linkcolor=blue]{hyperref}
\usepackage{authblk}

\usepackage[backend=bibtex,
			style=numeric,
			isbn=false,
			eprint=true,
			doi=false,
			url=false,
			giveninits=true,
			maxnames=4,
			sorting=none]{biblatex}
\renewbibmacro{in:}{}

\title{A Complete Invariant Analysis of the Kerr Spacetime and its Photon Region }
\author[1]{Nicholas T. Layden}
\author[2]{Dipanjan Dey}
\author[3]{Alan A. Coley}

\affil[1,3]{Department of Mathematics and Statistics, Dalhousie University, Halifax, Nova Scotia, Canada B3H 3J5}
\affil[2]{Beijing Institute of Mathematical Sciences and Applications, Beijing 101408, China}

\date{}

\begin{document}
	\maketitle
	
	\begin{abstract}
		We present an invariant characterization of the Kerr spacetime, and utilize the invariant structure of the spacetime to derive a function whose zeros identify a special family of null geodesics. Each member of this family is tangent to every photon surface in the Kerr photon region, offering a method of invariantly characterizing photon surfaces in axially symmetric spacetimes and thereby a providing a computational tool for efficiently computing the geodesic equations for any part of the photon region. The invariant that identifies all of the spherical photon orbits is parameterized by a Lorentz parameter, where the parameter is effectively an inclination angle of the spherical photon orbits through the equatorial plane. We also show how the invariant determines the constants of motion for all spherical orbits in the photon region. Finally, we briefly derive invariants which identify the other geometrically important surfaces such as the ergosurfaces and local horizons.
	\end{abstract}
	
	\section{Introduction}
	The Kerr solution in general relativity is considered one of the most plausible end states of gravitational collapse which forms a black hole \cite{Wald_1971}. The horizons, ergosurfaces, photon region, and shadow of the Kerr spacetime are of great importance when observing black holes in nature, and understanding these properties of a black hole is necessary to understand and test the properties of the universe  \cite{M87bh}. In modern observations of black holes in distant galaxies (and the one in our own galaxy), the black hole and its exterior properties are modeled principally using the Kerr or Kerr-Newman solutions. Mathematical invariants provide a tool for investigating properties of spacetimes that are truly invariant; that is, not observer or foliation-dependent (like apparent horizons), are locally defined (unlike event horizons), and detail all of the geometric properties of a spacetime. We approach the characterization of the Kerr photon regions in terms of the invariant structure of the spacetime by utilizing the Cartan-Karlhede algorithm. Among other things we provide a characterization of the Kerr event horizon in terms of invariants, which was described briefly in \cite{McNutt:2017paq}. Moreover, we show how the definition of a photon surface given by \cite{Claudel:2000yi} is enough to characterize the photon region in the Kerr spacetime, and generally motivates applications to other axially symmetric spacetimes.
	
	The Kerr spacetime is described by the following line element:
	
	\begin{dmath}\label{eq:kerr}
		ds^2= -\frac{Q-a^2\sin^2(\theta)}{R^2}dt^2-\frac{2a\sin^2(\theta)(r^2+a^2-Q)}{R^2}dtd\phi+\frac{\Bigr((r^2+a^2)^2-Qa^2\sin^2(\theta)\Bigr)\sin^2(\theta)}{R^2}d\phi^2 +\frac{R^2}{Q^2}dr^2+R^2d\theta^2,
	\end{dmath}
	where the functions $Q,R$ are defined by,
	
	\begin{equation}
		R\equiv \sqrt{r^2+a^2\cos^2(\theta)},\ \ Q\equiv r^2+a^2-2Mr,
	\end{equation}
	and the constant parameters $M$ and $a$ are the mass and spin of the central black hole, respectively. All quantities are in units with $G=c=1$.

	The photon region is of observational and theoretical importance in several ways. When viewed by an asymptotic observer, it has a characteristic appearance based on the mass and spin of the black hole, and therefore understanding the structure of the photon region allows us to potentially determine the properties of black holes we observe in the universe \cite{Ali_2025,EHT2019, Akiyama:2019fyp}. Moreover, the photon orbits around a black hole are closely connected to the behaviour of gravitational radiation via their quasi-normal mode peturbations \cite{Ferrari1984,Seidel1990}, and thus reveal the structure of the horizon and the behaviour of gravitational waves during mergers and black hole formation \cite{Kelly_2025}. The photon surface characteristics also determine the optical appearance of the black hole shadow, and hence the observable properties of the black hole \cite{Kumar_2020}.
	
	Past approaches characterizing the Kerr spacetime involve scalar polynomial invariants, usually constructed to identify the horizons and ergosurfaces invariantly by roots of some function of the polynomial invariants \cite{Abdelqader_2015}. We will approach the characterization of special surfaces instead by utilizing the Cartan scalars of the spacetime, which are defined below.

		In the Kerr spacetime, the event horizon and Cauchy horizon are the black hole surfaces (the event horizon considered the `physical' horizon), which are also Killing horizons. That is, the timelike killing vector $\xi^a$ has vanishing norm on these surfaces, $\xi^a\xi_a=0$. In general, a spacetime does not admit a globally defined horizon equivalent to the ones here, and thus local definitions are of much more use computationally. There are a handful of methods defining a horizon locally. The standard treatment is to consider the local expansion of a congruence of null geodesics, and the horizon itself is related to the vanishing of the expansion of outgoing/ingoing null geodesics. The surfaces identified by these conditions are referred to as apparent horizons, or more correctly marginally outer trapping surfaces \cite{Booth:2005qc}. An alternative approach conjectures that a horizon can be locally defined by the vanishing of some scalar polynomial invariant, or equivalently, a Cartan invariant, which is connected to the changing algebraic type of the horizon in comparison to the surrounding spacetime. The geometric horizon conjectures (in terms of Cartan invariants) \cite{Coley_2017,Coley_2017b} state that the horizon is defined locally by the vanishing of the highest (or lowest) boost weight components of the curvature tensor (for an algebraically general spacetime) or its derivatives (for an algebraically special spacetime). In the Kerr case the spacetime is Petrov Type D, which is algebraically special, therefore the geometric horizon is defined by the vanishing of the highest boost weight components of the derivative of the curvature tensor. Using our derivation of the invariant characterization of the curavture tensor and its derivatives by the boost weight decomposition, we derive a function which identifies the horizon \cite{McNutt:2017paq}.

	\subsection{Terminology and Definitions}
	
	A \textit{Newman Penrose null frame} is a null frame $\{ \ell,n,m,\bar{m}\}$ where the basis vectors satisfy the following conditions:
	
	\begin{subequations}
		\begin{align}
			\ell^a\ell_a = n^an_a = m^am_a = \bar{m}^a\bar{m}_a =0,\\ 
			-\ell^an_a = m^a\bar{m}_a =1,\\ 
			\text{all other contractions vanish.}
		\end{align}
	\end{subequations}
	In addition, the vectors $m,\bar{m}$ are complex conjugates.

	A particular form of a tensor defined in an invariant fashion with respect to a frame is called a \textit{canonical form}. Petrov Type D spacetimes all have the same Weyl tensor structure (based on principal null directions/eigenvalue structure). These are defined without reference to coordinates. There are `standard' canonical forms used in the literature, like the Petrov classification or the Segre classification \cite{joly}, but canonical forms can be defined in other ways. The canonical form structures we use in this work utilizes the alignment classification \cite{Coley:2004jv}, which is equivalent to the Petrov classification of the Weyl tensor and the Segre classification of the Ricci tensor.

	Given a set of $n$ functions of coordinates $\{ f_i(x^a)\}_{i=1}^n$, the functions are called \textit{functionally dependent} if $\exists \ F$ s.t. $F(f_1,\hdots,f_n)=0$, else they are independent. In general, the expression $df_1\wedge \hdots \wedge df_n \ne 0$ when $f_1,\hdots,f_n$ are \textit{functionally independent}.

	Under a Lorentz boost of the null frame, the \textit{boost weight} (bw) is defined as the power of the boost parameter a scalar acquires. Scalars defined by contracting with $\ell$ pick up a cumulative boost weight of $+1$, scalars defined by contracting with $n$ pick up a cumulative boost weight of $-1$, and contractions with $m$, $\bar{m}$ have cumulative boost weight $0$. The total boost weight is the sum of all of these contractions. The form of all Lorentz transformations is given below for the null frame.

	A frame is called \textit{Invariant(-ly defined)} if it is defined by the structure of the curvature tensors and its derivatives in a canonical form. This is derived via the Cartan-Karlhede algorithm.

	 All of the functionally independent components of the curvature tensor and its derivatives in the invariant frame are called \textit{Cartan scalars}. Frame derivatives of Cartan scalars are also Cartan scalars themselves. In $N$ dimensions, there are as many as $N$ functionally independent Cartan scalars.

	 Any scalar function which can be written as any combination of the Cartan scalars is called an \textit{Extended Cartan scalar}. The spin coefficients are usually such functions.

	 All (extended) Cartan scalars (as any scalar) are invariant under local coordinate transformations, and after the invariant frame has been fixed by the Cartan-Karlhede algorithm they are referred to as \textit{Cartan Invariants}. They are \textit{not} invariant under Lorentz transformations, but their form is determined by the structure of the curvature tensor and its derivatives.

	 Any scalar resulting from contractions of curvature tensors and their derivatives is called a \textit{Scalar Polynomial Invariant}. These functions are invariant under Lorentz transformations and local coordinate transformations. The Ricci scalar and Kretschmann scalar are frequently encountered examples of these.

	\subsection{Newman-Penrose Formalism}

	In the following analysis, we utilize the Newman-Penrose (NP) formalism. See \cite{Kramer} for a review of the formalism in general, and explicit definitions of the NP functions, and the appendix for some useful transformation rules. Here we briefly mention the functions defined in the formalism. Given an NP null frame (or tetrad), $\{ \ell,n,m,\bar{m}\}$, the components of the curvature tensor are redefined as a set of 15 real and complex functions, labelled by $\Psi_i$ ($i \in 0,1,2,3$), $\Phi_{ij}$ ($i,j \in 0,1,2$) and $\Lambda$, which are the components of the Weyl and traceless Ricci tensor relative to  the frame and the Ricci scalar, these encode all 20 real independent components of the Riemann tensor in 4 dimensions. The connection is a spin connection, whose components are written as a set of 12 complex functions, labelled by the greek letters, and are all directional derivatives of the frame basis vectors:
	\begin{equation}
		\alpha,\beta,\nu,\lambda,\rho,\mu,\epsilon,\gamma,\tau,\pi,\sigma,\kappa.
	\end{equation}
	We also define the frame derivatives as the directional derivatives:
	\begin{subequations}
		\begin{align}
			D \equiv \ell^a \nabla_a, \ \ \Delta \equiv n^a\nabla_a, \\
			\delta \equiv m^a \nabla_a, \ \ \bar{\delta} \equiv \bar{m}^a \nabla_a.
		\end{align}
	\end{subequations}

	\subsection{Lorentz Transformations of a NP Null Frame}
	
	For a Newman-Penrose null frame, there are 6 Lorentz parameters expressed through three classes of transformation of the null frame. These transformations are as follows:
	
	\noindent
	Null rotations about $l$:
	\begin{subequations}
		\begin{align}
			{\ell'}^a &=\ell^a, \\
			{n'}^a &= n^a + Em^a +\bar{E} \bar{m}^a+E\bar{E}\ell^a, \\
			{m'}^a &= m^a +\bar{E}\ell^a.
		\end{align}
	\end{subequations}
	Null rotations about $n$:
	\begin{subequations}
		\begin{align}
			{\ell'}^a &= \ell^a + \bar{B}m^a +B \bar{m}^a+B\bar{B}n^a, \\
			{n'}^a &=n^a, \\
			{m'}^a &= m^a +Bn^a,
		\end{align}
	\end{subequations}
	where the null rotation parameters $E,B \in \mathbb{C}$ are functions of the coordinates. The remaining transformations are spins and boosts, expressed together as a `spin-boost':
	\begin{subequations}
		\begin{align}
			{\ell'}^a &= A\ell^a, \\
			{n'}^a &= \frac{n^a}{A}, \\
			{m'}^a &= e^{i\Theta}m^a.
		\end{align}
	\end{subequations}
	The boost and spin parameters $A,\Theta \in \mathbb{R}$ with $A>0$, are also functions of the coordinates.

	\subsection{Numerically Solving the Geodesic Equations}
	Given a spacetime $(M,g)$, the motion of test particles in the spacetime is described by the geodesic equations. Suppose a test particle has position $x^a(\tau)$, parameterized by $\tau$, where $\tau$ is a suitable (possibly affine) parameter for the particle. The geodesic equations are:
	\begin{equation}
		\frac{d^2x^a}{d\tau^2} + \Gamma^a_{\ bc}\frac{dx^b}{d\tau}\frac{dx^c}{d\tau}=0,
	\end{equation}
	where $\Gamma^a_{\ bc}$ are the Christoffel symbols of the Levi-Civita connection for $g$. Solving these equations amounts two solving four coupled second order ordinary differential equations for $x^a(\tau)$. In general, we would not expect exact solutions, but in some cases of course this is possible. Suppose the test particle has a velocity vector given by $u^a(\tau)$, which has component functions:
	
	\begin{equation}
		u^a = \{u^1(\tau),u^2(\tau),u^3(\tau),u^4(\tau)\}.
	\end{equation}
	The particle is either timelike, null, or spacelike when the inner product $u^au_a$ is negative, zero, or positive, respectively. The character of the particle in question is then determined by defining proper initial data which satisfies the timelike null or spacelike conditions. The position of the particle is given by the vector,
	
	\begin{equation}
		x^a(\tau) = \{x^1(\tau), x^2(\tau), x^3(\tau), x^4(\tau) \}.
	\end{equation}
	To tailor the geodesic equations to a more traditional numerical context, we can define the variables,
	
	\begin{subequations}
		\begin{align}
			u^1(\tau) = \frac{dx^1}{d\tau},\ 
			u^2(\tau) = \frac{dx^2}{d\tau},\\
			u^3(\tau) = \frac{dx^3}{d\tau},\ 
			u^4(\tau) = \frac{dx^4}{d\tau}.
		\end{align}
	\end{subequations}
	Substituting these new variables into the geodesic equations above gives us the eight dimensional system of first order differential equations:
	
	\begin{subequations}
		\begin{align}
			\frac{dx^a}{d\tau} &= u^a,	\\
			\frac{du^a}{d\tau} &= - \Gamma^a_{\ bc}u^b u^c.
		\end{align}
	\end{subequations}
	These equations can then be solved as an initial value problem by specifying the initial data $\{x^a(0),u^a(0)\}$. 
	
	It is important to note that when applied to the Kerr spacetimes (or other highly symmetric spacetimes), it is much more advantageous to solve the geodesic equations by finding their first integrals, and using the constants of motion to describe the orbits. We utilize both approaches in this work to show how the applications of the invariant characterization offer a method of analyzing the structure of the geodesics in the spacetime. See Section \ref{ch:firstintegralequations} for this analysis.

	\subsection{Invariant Structure of Spacetimes}
	The way we approach deriving invariant properties of a spacetime (or any differentiable manifold in general) is by the Cartan-Karlhede algorithm \cite{Kramer}. It turns out that scalar polynomial invariants (SPIs) (any contractions of the curvature tensors and/or their derivatives of any order), which are scalar invariants of a spacetime, do not fully characterize a spacetime \cite{Coley:2004jv}. The most obvious way to see this is to observe that the Minkowski and the degenerate K\"undt spacetimes, which describe empty space and propagating gravitational waves respectively both have all of their SPIs vanishing everywhere. So by analyzing the SPIs alone, in general you cannot distinguish between these two radically different spacetimes, and hence not all spacetimes can be characterized by their SPIs.
	
	 Cartan first presented a method to establish a method to find local equivalences between spacetimes, which was later enhanced by Karlhede, who showed that by using tetrads (frames) with constant components, one could in principal determine the local equivalence of two spacetimes with only at most seven derivatives of the Riemann tensor. In practice, less spacetime symmetries actually makes this process easier; in general, for a Petrov Type I spacetime the Cartan invariants can all be derived at zeroth order. As the number of symmetries (and their type) increases in a spacetime, generally the more derivatives of the Riemann tensor are required for its characterization. In the physically important cases like spherically symmetric, axially symmetric spacetimes (stationary or dynamic), the number of derivatives required is at most two, and usually only one. An interesting case is the Szekeres spacetime which, in general, has no symmetries at the level of the metric, yet an invariant characterization is completely derived by the first derivative of the Riemann tensor \cite{Coley:2019ylo}.

	\section{Invariant Characterization of the Kerr Spacetime}
	We use the Cartan-Karlhede algorithm to develop an invariant characterization of the spacetime. The starting point is an initial NP null frame, of which there are a few popular null frames used in the literature; for example, Kinnersley's tetrad \cite{Kinnersley1969} or the null frame given in \cite{McNutt:2017paq}. This initial choice does not matter, so we will choose to start with Kinnersley's tetrad $\{\ell_0,n_0,m_0,\bar{m}_0 \}$, given in components:
	
	\begin{subequations}
		\begin{align}
			\ell_0^a &= \left\{\frac{r^2+a^2}{Q},1,0,\frac{a}{Q} \right\}, \\
			n_0^a &= \left\{\frac{r^2+a^2}{2R^2},\frac{-Q}{2R^2},0,\frac{a}{2R^2} \right\}, \\
			m_0^a &= \left\{\frac{ia\sin(\theta)}{\sqrt{2}(ia\cos(\theta)+r)},\frac{1}{\sqrt{2}(ia\cos(\theta)+r)},0,\frac{i\csc(\theta)}{\sqrt{2}(ia\cos(\theta)+r)} \right\}, \\
			\bar{m_0}^a &= \left\{\frac{-ia\sin(\theta)}{\sqrt{2}(-ia\cos(\theta)+r)},\frac{1}{\sqrt{2}(-ia\cos(\theta)+r)},0,\frac{-i\csc(\theta)}{\sqrt{2}(-ia\cos(\theta)+r)} \right\}.
		\end{align}
	\end{subequations}
	By projecting the frame onto the curvature tensors, we proceed to determine the algebraically independent components. Since the Kerr solution is vacuum, the Ricci tensor and Ricci scalar are identically zero, and we only find the following components of the Weyl tensor to be in this initial frame,
	
	\begin{equation}\label{eq:psi2}
		\Psi_2 = \frac{iM}{(a\cos(\theta)+ir)^3}, \ \bar{\Psi}_2= \frac{iM}{(a\cos(\theta)+ir)^3}.
	\end{equation}
	All other non-zero components of the Weyl tensor are constant multiples of these functions. These functions are both boost and spin invariant, thus we still have to fix boosts and spins of the frame. The zeroth order of the Cartan-Karlhede algorithm thus concludes with two functions, and the isotropy group has been reduced from six dimensions to two dimensions. The two components above are functionally independent, by the simple computation:
	
	\begin{equation}
		d\Psi_2 \wedge d\bar{\Psi}_2=-\frac{18iM^2a\sin(\theta)}{(r^2+a^2\cos^2(\theta))^4}dr\wedge d\theta \ne 0.
	\end{equation}

	 We now proceed to first order, to fix the remaining Lorentz parameters. At first order we have the Weyl tensor which contains components proportional to $\Psi_2$, and its directional derivatives. The quantities appearing at first order are affected by boosts and spins, thus we can fix the boost and spin parameters by demanding that the functions among the derivative of the Weyl tensor satisfy,
	
	\begin{equation} \label{eq:cartan2}
		D\Psi_2=-\Delta \Psi_2, \ \delta \Psi_2 = -\bar{\delta} \Psi_2.
	\end{equation}
	Both of these conditions fix the boost and spin to be,
	
	\begin{subequations}
		\begin{align}
			A&=\sqrt{\frac{-\Delta \Psi_2}{D\Psi_2}}=\sqrt{\frac{a^2+r(r-2M)}{a^2+a^2\cos^2(\theta)+2r^2}},\\
			\exp(i\Theta) &=  \sqrt{\frac{-\bar{\delta}\Psi_2}{\delta\Psi_2}}=\sqrt{\frac{a\cos(\theta)+ir}{a\cos(\theta)-ir}}.
		\end{align}
	\end{subequations}
	The algorithm stops here, as all Lorentz parameters have been fixed, and the maximum number of functionally independent invariants has been derived. The two functionally independent invariants which replace the $r$ and $\theta$ coordinates can be given in a simpler form by combining the zeroth order invariants,
	
	\begin{equation}
		C_1 = \text{Re}\left\{\left(\frac{i}{\Psi_2}\right)^{1/3}\right\},\ C_2 = \text{Im}\left\{\left(\frac{i}{\Psi_2}\right)^{1/3}\right\}.
	\end{equation} 
	Once these invariants are derived, and replace the coordinates in all components of the curvature tensors, one can compare the resulting functions to another spacetime (possibly) related by a coordinate transformation, which is the essence of the original equivalence algorithm by Cartan.
	
	We now list the invariant frame vectors in components:
	
	\begin{subequations}
		\begin{align}
			\ell^a &= \left\{\frac{a^2+r^2}{\sqrt{2}\sqrt{Q}R},\frac{\sqrt{Q}}{\sqrt{2}R},0,\frac{a}{\sqrt{2}\sqrt{Q}R} \right\}, \\
			n^a &= \{\frac{a^2+r^2}{\sqrt{2}\sqrt{Q}R},-\frac{\sqrt{Q}}{\sqrt{2}R},0,\frac{a}{\sqrt{2}\sqrt{Q}R} \}, \\
			m^a &= \left\{\frac{a\sin(\theta)}{\sqrt{2}\sqrt{a^2\cos^2(\theta)+r^2}},0, -\frac{i}{\sqrt{2}\sqrt{a^2\cos^2(\theta)+r^2}},\frac{\csc(\theta)}{\sqrt{2}\sqrt{a^2\cos^2(\theta)+r^2}}\right\}, \\
			\bar{m}^a &= \left\{\frac{a\sin(\theta)}{\sqrt{2}\sqrt{a^2\cos^2(\theta)+r^2}},0 ,\frac{i}{\sqrt{2}\sqrt{a^2\cos^2(\theta)+r^2}},\frac{\csc(\theta)}{\sqrt{2}\sqrt{a^2\cos^2(\theta)+r^2}}\right\}.
		\end{align}
	\end{subequations}

	With the two Cartan invariants, all components of the curvature tensors of all orders can be described, as well as the fundamental invariant properties of the Kerr spacetime, the black hole spin parameter $a$ and the black hole mass $M$. The completely fixed frame is now referred to as the \textit{invariantly defined null frame}, and we now list the Newman-Penrose quantities in the invariant frame. The spin coefficients are given by

	\begin{subequations}
		\begin{align}
			\mu &= \rho = \frac{\sqrt{a^2+r(r-2M)}}{(ia\cos(\theta)+r)\sqrt{a^2+a^2\cos(2\theta)+2r^2}}, \\ 
			\gamma &= \epsilon = \sqrt{\frac{a^2+a^2\cos(2\theta)+2r^2}{a^2+r(r-2M)}}\left(\frac{\left(-a\cos(\theta)(M-r) - i(a^2-Mr) \right)}{4(ia\cos(\theta)-r)(ia\cos(\theta)+r)^2}\right), \\
			\pi &= \tau = -\frac{ia\sin(\theta)\sqrt{a\cos(\theta)-ir}}{\sqrt{2}(a^2\cos^2(\theta)+r^2)\sqrt{a\cos(\theta)+ir}}, \\
			\alpha &= \beta = \frac{a\csc(\theta)(3+\cos(2\theta))-2i\cot(\theta)(a^2-r^2)}{4\sqrt{2}(a\cos(\theta)+ir)^{5/2}\sqrt{a\cos(\theta)-ir}},\\
			\lambda &= \kappa = \sigma = \nu = 0.
		\end{align}
	\end{subequations}
	The only non-zero Ricci-NP and Weyl-NP scalar is $\Psi_2$, given above in (\ref{eq:psi2}). See Table \ref{tab:invariant} for the full invariant definition of the frame.
		
	\begin{table}[h!]
	\centering
	\caption{Construction of the invariant null frame for the Kerr spacetime. The number $q$ is the derivative order of the curvature tensor. The $t_q$ is the number of functionally independent components up to the $q$th order. $\dim H_q$ is the dimension of the isotropy group (number of free Lorentz transformations) at each $q$. The algorithm continues until $t_q$ and $\dim H_q$ stop changing, and no Lorentz freedom remains, which occurs for order $q=2$. }
		\begin{tabular}{c|c|c|c|c}
			Order $q$ &$t_q$ & $\dim H_q$ & Fixed Transformations & Canonical Choice \\ \hline
			0 & 2 & 2 & Null Rotations& Type D Weyl Tensor \\
			1 & 2 & 0 & Spins, Boosts & $D\Psi_2 = -\Delta \Psi_2$, \ $\delta \Psi_2 = -\bar{\delta}\Psi_2$ \\
			2 & 2 & 0 & None & Frame completely fixed
		\end{tabular}
		\label{tab:invariant}
	\end{table}
	In the invariant frame, it will also be useful to define the radial direction in terms of the invariant frame vectors:
	
	\begin{equation}
		\frac{1}{\sqrt{2}}(\ell-n)=\frac{\sqrt{Q}}{R}\partial_r.
	\end{equation}
	We also list the algebraically independent components of the derivative of the Weyl tensor, sorted by boost weight:
	
	\begin{subequations}
		\begin{align}
			\text{bw $-1$: }& -3(\Psi_2\rho + \bar{\Psi}_2\bar{\rho}), \ 3\Psi_2 \rho, \  3\bar{\Psi}_2 \bar{\rho}, \ 3\Psi_2 \rho - 3\bar{\Psi}_2 \bar{\rho}, \\
			\text{bw $0$: }& 3(\Psi_2\tau + \bar{\Psi}_2\bar{\tau}), \ 3\Psi_2 \tau, \  3\bar{\Psi}_2 \bar{\tau}, \ -3\Psi_2 \tau + 3\bar{\Psi}_2 \bar{\tau},\  \\
			\text{bw $+1$: }& 3(\Psi_2\rho + \bar{\Psi}_2\bar{\rho}), \ 3\bar{\Psi}_2\bar{\rho}, \ 3\Psi_2 \rho \ -3\Psi_2 \rho + 3\bar{\Psi}_2 \bar{\rho}.
		\end{align}
	\end{subequations}
	Clearly there are two unique sets of components, the relationship among these components identifies the horizon and ergosurfaces, as shown in the following.

	\subsection{Characterization of the Horizon}
	Here we utilize the geometric horizon conjecture to produce a Cartan invariant that detects both of the horizons. Using the Bianchi identities, it can be shown that in the invariant null frame, the following identity holds:
	
	\begin{equation}
		D\Psi_2 = 3\rho \Psi_2,
	\end{equation}
	and according to the boost weight classification of the spacetime, this is the highest boost weight component of the Kerr spacetime. It has been shown that black hole spacetimes are characterized by their curvature invariants \cite{Coley_2009}, where the horizons are more algebraically special than the exterior spacetime \cite{Coley_2017b}. The locus of the vanishing of the highest (or lowest) boost weight components identifies a geometrically special region. Notice that many of the Weyl tensor components at first order are proportional to the spin coefficient $\rho$, and when $\rho=0$, the structure of the first order Weyl tensor changes, which identifies the local horizon geometrically. In coordinates, the zeros of $\rho$ give the usual location of the horizons:
	
	\begin{equation}
		\rho = 0 \iff  r(r-2M)+a^2=0,
	\end{equation}
	where the outer and inner horizons (analogous to the event and Cauchy horizons respectively) are located at
	\begin{equation}
		r_{\pm} = M \pm \sqrt{M^2-a^2}.
	\end{equation}
	For an SPI which detects the horizons and gives an invariant characterization of the Kerr spacetime (in terms of SPIs) see \cite{Abdelqader_2015}.
	
	\subsection{Characterization of the Ergosurfaces}
	From the Bianchi identities, we find the following
	\begin{equation}
		\delta \Psi_2 = 3\tau\Psi_2.
	\end{equation}
	
	In terms of the curvature invariants, the square of the first derivative of the Riemann (Weyl) tensor is given in terms of the NP quantities in the invariant frame as,
	
	\begin{equation}
		C_{abcd;e}C^{abcd;e}=720\Bigr(\Psi_2^{\ 2}(\rho^2-\tau^2)  + \bar{\Psi}_2^{\ 2}(\bar{\rho}^2-\bar{\tau}^2) \Bigr),
	\end{equation}
	the functions $\rho^2 -\tau^2$ and $\bar{\rho}^2-\bar{\tau}^2$ both vanish when the following expression vanishes,
	
	\begin{equation}
	 	a^2\cos^2(\theta) + 2r(r-2M)=0,
	\end{equation}
	identifying the ergosurfaces where the extended Cartan invariant $\rho^2-\tau^2=0$, or equivalently where the scalar polynomial invariant $C_{abcd;e}C^{abcd;e}=0$. The explicit expressions for the ergosurfaces in these coordinates are:
	
	\begin{equation}
		r_{\rm ergo\pm } = M \pm \sqrt{M^2-a^2\cos^2(\theta)}.
	\end{equation}
	In the structure of the first derivative of the Weyl tensor, the ergosurface is identified in the invariant frame where all components of the Riemann tensor are proportional to the functions $\text{Re}(\rho),\text{Im}(\rho),\rho,\bar{\rho}$.

	\subsection{Characterization of the Photon Region}
	
	\textbf{Definition: (Photon Surface)}
	
	\noindent
	\textit{
	Let $(\mathcal{M},g)$ be a spacetime. A photon surface of $(\mathcal{M},g)$ is an immersed, nowhere spacelike hypersurface $S\subset M$ such that for every point $p\in S$, for all null vectors there exists a null geodesic $\gamma:(-\epsilon,\epsilon)\to M$ such that $\dot{\gamma}(0)=k_{i}, \ |\gamma|\subset S$.}

	To examine the photon region, we need to utilize the invariant frame to find null geodesic vectors which satisfy the photon surface conditions. A timelike photon surface $\mathcal{S}$ is a surface identified by null tangent vectors $\ell^a$ and a spacelike normal vector $N^a$, where $\ell^a$ is null geodesic on $S$, and is null geodesic in the full spacetime, $\mathcal{M}$. From our invariant characterization, we have the vector $\ell^a$ which happens to be null geodesic everywhere in the spacetime ($\kappa=0$), but is not necessarily affinely parameterized, since $\epsilon+\bar{\epsilon}\ne0$ in general. What we aim to do here is to provide a set of Lorentz transformations which aligns the null vector $\ell^a$  to every possible photon surface in the spacetime.
	
	Past investigations of photon orbits and photon surfaces in the Kerr solution (and general stationary axisymmetric electrovacuum) \cite{Paganini2016,Grenzebach_photon_regions}, have shown that a photon surface exists in the spacetime only if it has spherical topology, and based on these past analyses, we can write down the radial spacelike normal as:
	
	\begin{equation}
		N^a = \left\{ 0,\frac{\sqrt{Q}}{R},0,0\right\}.
	\end{equation}
	A photon surface is characterized by a normal $N^a$, and a null geodesic vector $\ell^a$ which is null geodesic in the spacetime (and on the photon surface), and satisfies the photon surface condition
	
	\begin{equation}\label{eq:pscond}
		\ell^a \ell^b \nabla_a N_b =0.
	\end{equation}
	By choosing the alignment of the null vectors $\ell^a$ such that they are tangent to a photon surface, we can show that on each photon surface identified in the region by the proceeding analysis, this expression vanishes. In addition to identifying the timelike photon surfaces that foliate the photon region, this also identifies the two horizons, since they are null hypersurfaces, and thus trivially photon surfaces themselves. The normal we define above happens to become null on the horizons, based on the definition we give in 

	Since the spacetime is stationary, choosing the form of the normal for the photon surfaces is rather straightforward, whereas in the dynamic cases the situation is much more technical and requires more analysis (see \cite{Dey:2024vne} for an exploration of the photon surfaces in dynamic spherically symmetric spacetimes).
	
%
%
%

\subsection{Photon Surface Frame}

	Any null frame which is aligned to the null tangents of a photon surface will be referred to as a `photon frame'. Photon frames will not be unique, as there are other frames related by Lorentz transformations, that also identify the photon surfaces, and we aren't deriving them from the geometry directly. Photon surfaces also do not appear to be identified directly by the components of the Riemann tensor or its derivatives in the invariant frame, rather we will find that there is a parametric function which identifies them all in this spacetime.

	Consider the frame $\{\tilde{\ell}^a,\tilde{n}^a,\tilde{m}^a,\tilde{\bar{m}}^a \}$ which is related to the invariant frame by a null rotation about $n^a$ by $B=e^{iK}$ ($K\in \mathbb{R}$), the vector $\tilde{\ell}^a$ is:
	
	\begin{equation}
		\tilde{\ell}^a=\ell^a + e^{-iK}m^a + e^{iK} \bar{m}^a + n^a.
	\end{equation}
	We can also define the normal $N^a$ directly in terms of null frame vectors:
	\begin{equation}
		N^a=\frac{1}{\sqrt{2}}(\ell^a-n^a).
	\end{equation}
	The photon surface condition written in terms of the invariant frame spin coefficients is then given explicitly by
	\begin{equation}\label{eq:photon}
		\tilde{\ell}^a\tilde{\ell}^bK_{ab}=\frac{1}{\sqrt{2}}\tilde{\ell}^a\tilde{\ell}^b\nabla_a(\ell^a-n^a)=\frac{1}{\sqrt{2}}\left[2(\epsilon + \bar{\epsilon}) + \rho + \bar{\rho} +\cos(K)(2\beta+2\bar{\beta} +\tau + \bar{\tau})\right],
	\end{equation}
	which can now be expanded in Boyer-Lindquist coordinates as
	
	\begin{equation}\label{eq:ps-bl}
		P(K,r,\theta)\equiv\tilde{\ell}^a\tilde{\ell}^bK_{ab} = \frac{\sqrt{Q}\Bigr(a^2(r+3M)+r^2(r-3M) +a^2\cos(2\theta)(M-r)\Bigr) + 4arQ\sin(\theta)\cos(K)}{(a^2\cos^2(\theta)+r^2)Q^{3/2}}.
	\end{equation}
	This function is now an expression which identifies all photon surfaces in the photon region, and is parameterized by the continuous parameter $K$. To prove that this identifies a photon surface, we examine the null geodesic equations  for the vector field $\tilde{\ell}^a$.

When compared to the static spherically symmetric spacetimes \cite{Dey:2024vne}, it is interesting to note that the expression in (\ref{eq:photon}) is similar in form. The extra terms multiplied to the $\cos K$ term do not appear, and the invariant photon surface expression is $(2\epsilon+\rho)=0$. In an axisymmetric spacetime, clearly the effect of asymmetry introduces some complications to the structure of the photon surface itself. Indeed, some others approach identifying photon surfaces in axial symmetry (stationary as well as non-rotating), via different geometric approaches \cite{Kobialko}. In most other approaches, a modification of the definition of a photon surface seems necessary.

	\subsection{Constant Null Rotations of the Invariant Frame}
	
	The process here is essentially as follows. We wish to transform the invariant null frame so that it is aligned to the null geodesics of the spacetime, but not such that the frame is aligned to null geodesics \textit{everywhere}, rather only when the frame is on a photon surface.  It is important to note that we do in fact have a frame which is null geodesic everywhere, which is the Kinnersley frame defined above. This frame is not sufficient to detecting the photon surfaces, and similarly, the invariant null frame on its own also does not seem to be able to detect the photon surfaces. There is a possibility that like the horizon and ergosurfaces, the photon surface can be identified by certain scalar polynomial invariants, but we do not take that approach here. Rather, the frames themselves will never have the real null vectors tangent to a timelike photon surface. 

	For a general NP frame $\{\ell^a,n^a,m^a,\bar{m}^a\}$, the geodesic equations for $\ell^a$ can be expanded in terms of the null frame and spin coefficients in the following way \cite{Kramer}:
	
	\begin{equation}
		D\ell^a = \ell^a \nabla_a \ell^b = (\epsilon + \bar{\epsilon})\ell^b - \bar{\kappa} m^b -\kappa\bar{m}^b.
	\end{equation}
	When $\kappa=0$, the null vector $\ell^a$ is geodesic, and further, when $\epsilon+\bar{\epsilon}=0$, it is affinely parameterized. This means that the character of the vector $\ell^a$ is determined by the value of the spin coefficients $\kappa,\epsilon+\bar{\epsilon}$ in the null frame. Suppose we transform the frame by a null rotation about $n^a$ with parameter $B\in \mathbb{C}$, the spin coefficient expressions $\kappa$ and $\epsilon +\bar{\epsilon}$ in a general frame become:
	
	\begin{subequations}
		\begin{align}
			\tilde{\epsilon} + \bar{\tilde{\epsilon}} &= \epsilon + \bar{\epsilon}+B^2( \lambda +\bar{B}\nu) + B(\alpha + \bar{\beta} + \pi + \bar{B} (\gamma + \bar{\gamma} + \mu + \bar{\mu} + \bar{B}\bar{\nu})) + \bar{B}(\bar{\alpha}+\beta + \bar{\pi} +\bar{B}\bar{\lambda}), \\
			\tilde{\kappa} &= \kappa + B(2\epsilon +\rho) + \bar{B}\sigma + B\bar{B}(2\beta+ \tau) + B^2(2\alpha + \pi) + B^3\lambda  + B^2\bar{B}(2\gamma +\mu) + B^3\bar{B}\nu  \\ 
			&\ \ \ -DB-B\bar{B}\Delta B - \bar{B}\delta B-B\bar{\delta}B.\nonumber
		\end{align}
	\end{subequations}
	The tilde denotes the quantity in the transformed frame $\{\tilde{\ell}^a, \tilde{n}^a, \tilde{m}^a, \bar{\tilde{m}}^a \}$. If we started in the invariantly defined null frame for Kerr, we can simplify these expressions greatly down to
	
	\begin{subequations}
		\begin{align}
			\tilde{\epsilon} + \bar{\tilde{\epsilon}} &= \epsilon + \bar{\epsilon} + B(\beta + \bar{\beta} +\tau) + \bar{B}(\beta + \bar{\beta} + \bar{\tau}) + B\bar{B}(\epsilon + \bar{\epsilon} + \rho + \bar{\rho}),\\
			\tilde{\kappa} &= 2\epsilon + \rho+ (B+\bar{B})(2\beta + \tau) + B\bar{B}(2\epsilon + \rho) -DB-B\bar{B}\Delta B - \bar{B}\delta B-B\bar{\delta}B.
		\end{align}
	\end{subequations}
	Since we want to locate points in the spacetime where $\tilde{\ell}^a$ is tangent to a null geodesic curve, let us propose the following. Suppose $B=\exp(iK)$, where $K\in [0,2\pi]$ is a constant. The transformed spin coefficients reduce
	
	\begin{subequations}
		\begin{align}
			\tilde{\epsilon} + \bar{\tilde{\epsilon}} &= (2\epsilon + 2\bar{\epsilon} + \rho + \bar{\rho}) + \exp(iK)(\beta + \bar{\beta} +\tau) +\exp(-iK)(\beta + \bar{\beta} + \bar{\tau}) ,\\
			\tilde{\kappa} &= 2\exp(iK)\left[2\epsilon + \rho+ \cos(K)(2\beta + \tau)\right]. \label{eq:kappatilde}
		\end{align}
	\end{subequations}
	By specifying the phase $K$, we can find where both expressions vanish. An easy way to start is let $\theta = \pi/2$, that is, we are aligning the null vector to null geodesics emanating from the equatorial plane. Then specifying a value for $K$ and solving for $r$ where $\kappa = \epsilon+\bar{\epsilon}=0$ gives us a family of null geodesic vectors parameterized by $K$, which are null geodesic at different points $(r,\pi/2,\phi)$ in the spacetime. Then evolving the geodesic equations away from this initial point will trace the photon surfaces. For each $K$, a surface of radius $r=r_K$ is identified by the vanishing of the spin coefficients above. As we will see, the `surfaces' here are not complete spheres, but ones which have been sheared off at that top and bottom, at different heights for different radii.
	
		For the null rotation above, in the equatorial plane, both expressions evaluate to
		
		\begin{equation}
			\kappa,\ \epsilon+\bar{\epsilon} \propto 2a^2 + r(r-3M) +2a\cos(K)\sqrt{a^2+r(r-2M)},
		\end{equation}
	where the values $r$ where this function vanishes for each value of the parameter $K \in [0,2\pi]$. The parameter $K$ identifies each spherical photon surface in the family of the surfaces in the Kerr photon region. Each of these surfaces is known to be individually unstable, that is, moving a null geodesic vector slightly off its particular photon surface will make the null geodesic either fall into the black hole, or escape to infinity. The `special' photon orbits that remain in the equatorial plane correspond to $K=0,\pi$, which are roots of a fourth order polynomial. There is a simple solution when $K=\pi/2$, which reduces to a quadratic, and has roots given by
	
	\begin{equation}
		r_{K=\pi/2}=\frac{3M}{2}\pm \frac{\sqrt{9M^2-8a^2}}{2},
	\end{equation}
	where the larger root is one of the outer photon surfaces, and the smaller is another inside the horizon. Both of the roots in this expression locate the only complete spheres inside the photon region.
	
	In the limit $\theta\to 0$, the function $\tilde{\epsilon} + \bar{\tilde{\epsilon}}$ reduces to 
	
	\begin{equation}
		\tilde{\epsilon} + \bar{\tilde{\epsilon}}|_{\theta=0}=-\sqrt{2}\frac{(Ma^2+a^2r-3Mr^2+r^3)}{(a^2+r^2)^{3/2}\sqrt{a^2+r(r-2M)}},
	\end{equation}
	which has a real root at the intersection point of the boundary of the photon regions, which gives the radius of the only sphere which is entirely contained within the photon region.

	\subsection{Variable Null Rotation}
	If instead of assuming that $K$ is constant, we suppose that $K=K(r)$, then the expression for $\tilde{\kappa}$ reduces to the same result as in equation (\ref{eq:kappatilde}), as the derivative portions of the transformation are identically zero for arbitrary $K(r)$. In this case, we can solve for $\tilde{\kappa}|_{\theta=\pi/2}=0$ for $\cos(K(r))$, or specify $K(r)$ such that for each $K$, the null vector $\ell^a$ is aligned to null geodesics in the equatorial plane, and hence the orbits which trace the photon surfaces. Since $\tilde{\kappa}$ depends on $\theta$, the $K(r)$ specified will only align the null vectors $\ell^a$ for a specified $\theta$. If we choose $\theta=\pi/2$, then the null vectors can be chosen to be aligned to null geodesics (and the photon surfaces) for every $r$ in the equatorial plane of the photon region.
	
	Expanding $\tilde{\kappa	}$ in terms of spin coefficients results in
	\begin{equation}
		\tilde{\kappa}= 2e^{i K}\Bigr( 2\epsilon + \rho + \cos(K)\left(2\beta + \tau\right)\Bigr),
	\end{equation}
	and in terms of coordinates,
	
	\begin{equation}\label{eq:cosk}
		\tilde{\kappa}= \frac{e^{iK}}{\sqrt{2}\sqrt{Q}R}\left[\mathcal{K}\right],
	\end{equation}
	where $\text{Re}(\mathcal{K}), \text{Im}(\mathcal{K})$ are the real and imaginary terms inside the brackets respectively, and are given by
	
	\begin{subequations}
		\begin{align}
			\text{Re}(\mathcal{K}) &= 2r\sqrt{Q}\cos(K)\cot(\theta)+2a\cos(\theta)(r-M)R, \\
			\text{Im}(\mathcal{K}) &= \sqrt{Q}a\cos(K)(\cos(2\theta)-3)\csc(\theta)+2(rM-a^2-Q)R.
		\end{align}
	\end{subequations}
	Expanding the expression for $\tilde{\epsilon}+\bar{\tilde{\epsilon}}$ gives,
	
	\begin{dmath}
		\tilde{\epsilon} + \bar{\tilde{\epsilon}} = \frac{1}{\sqrt{2} \sqrt{Q} R}\left[ 2 a  \sin (\theta ) \sqrt{Q}
   R (a  \cos (\theta ) \sin (K)-2
   r  \cos (K))+a ^2 \cos (2 \theta ) (r -M ) R^2-\left(M  \left(a ^2-6 r ^2\right)+3 r  a ^2+2
   r ^3\right) R^2\right].
	\end{dmath}
	The real roots $\{(r,\theta)\ |\ \tilde{\epsilon} + \bar{\tilde{\epsilon}}=0,\tilde{\kappa}=0\}|_K$ identify every photon surface in the entire photon region traced by affinely parameterized null geodesics tangent to $\tilde{\ell}^a$. In general, solving this equation is a numerical problem, but as shown above, for a few choices of $\theta$, it can reduce to a simple expression that can be solved analytically.

	\subsection{Spin transformation of the frame}
	Having noticed the behaviour of the spin coefficients $\kappa$ and $\epsilon$ under null rotations, it may provide some more insight to perform another Lorentz transformation of the frame. In particular, consider performing a spin transformation before the null rotation. Then the spin coefficients in question transform under a spin in the following way
	
	\begin{equation}
		\tilde{\tilde{\kappa}} = e^{i\Theta}\kappa, \ \ \tilde{\tilde{\epsilon}} + \bar{\tilde{\tilde{\epsilon}}}=\epsilon + \bar{\epsilon},
	\end{equation}
	clearly the real part of $\epsilon$ is unaffected by the spin, and $\kappa$ simply picks up a factor of $e^{i\Theta}$. If we perform a spin by the parameter $\Theta=-K$, and a null rotation as in the previous section, these spin coefficients transform to
	
	\begin{subequations}
		\begin{align}
			\tilde{\tilde{\kappa}} &= 2\Bigr( 2\epsilon + \rho + \cos(K)\left(2\beta + \tau\right)\Bigr), \\
			\tilde{\tilde{\epsilon}} + \bar{\tilde{\tilde{\epsilon}}} &= (2\epsilon + 2\bar{\epsilon} + \rho + \bar{\rho}) + \exp(iK)(\beta + \bar{\beta} +\tau) +\exp(-iK)(\beta + \bar{\beta} + \bar{\tau}).
		\end{align}
	\end{subequations}
	The photon surface condition can also be expressed in terms of the spin coefficients similarly:
	
	\begin{equation}
		\tilde{\ell}^a\tilde{\ell}^bK_{ab} =  -\sqrt{2}\left[2\epsilon + 2\bar{\epsilon} + \rho + \bar{\rho} + \cos(K)\left(2\beta + 2\bar{\beta} + \tau + \bar{\tau}\right)\right] \propto \tilde{\tilde{\kappa}}+\bar{\tilde{\tilde{\kappa}}}.
	\end{equation} 
	Since geodicity of $\tilde{\ell}^a$ requires at least $\tilde{\tilde{\kappa}}=\bar{\tilde{\tilde{\kappa}}}=0$, and we immediately recognize the photon surface condition is proportional to the real part of $\tilde{\tilde{\kappa}}$, then in this frame, the vector field $\tilde{\ell}^a$ is precisely null geodesic if and only if it is on a photon surface. 
	
	Approaching the Lorentz transformations in this way, and examining the geodesic conditions for $\tilde{\ell}^a$ produces the same conditions as derived directly from the photon surface conditions, resulting in the invariant $P$. In a more general (non-symmetric) spacetime, we would not expect that the normal to the photon surface can be derived as easily in the Kerr case from the invariant frame vectors. Thus a deeper exploration into the geodesic equations, and the Lorentz transformations of the frame are necessary. Indeed, this approach is utilized in \cite{Dey:2024vne}, and in (currently unpublished) results in the Szekeres spacetime by the present authors.

	\section{Behaviour of the Null Geodesics on the Photon Surfaces}
	We see some interesting behaviour of the null geodesic tangents $\tilde{\ell}^a$ emanating from the equatorial plane as the invariant frame vector $\ell^a$ is null rotated by a phase angle $K$. With no spin ($a=0$), any null rotation by the phase $K$ identifies the Schwarzschild photon surface $r=3M$. As the spin increases, what happens is the null geodesic vectors with different phase angles separate out into different distinct photon spheres with different radii, the vector with $K=\pi$ identifies the outermost photon ``sphere'', which actually is the outer photon orbit in the equatorial plane, for $K=0$ the innermost photon ``sphere'' is the inner photon orbit in the equatorial plane, identified by the roots of the polynomial
	
	\begin{equation}
		r^3 -6Mr^2+9M^2r-4Ma^2=0.
	\end{equation}
	The smallest root of the polynomial is always inside the two horizons. With the phase angle $K=\pi/2$, as mentioned above, the expression for the identification of the photon surfaces simplifies to a quadratic function
	
	\begin{equation}
		r(r-3M)+2a^2=0.
	\end{equation}

	\begin{figure}
			\centering
			\includegraphics[scale=0.04]{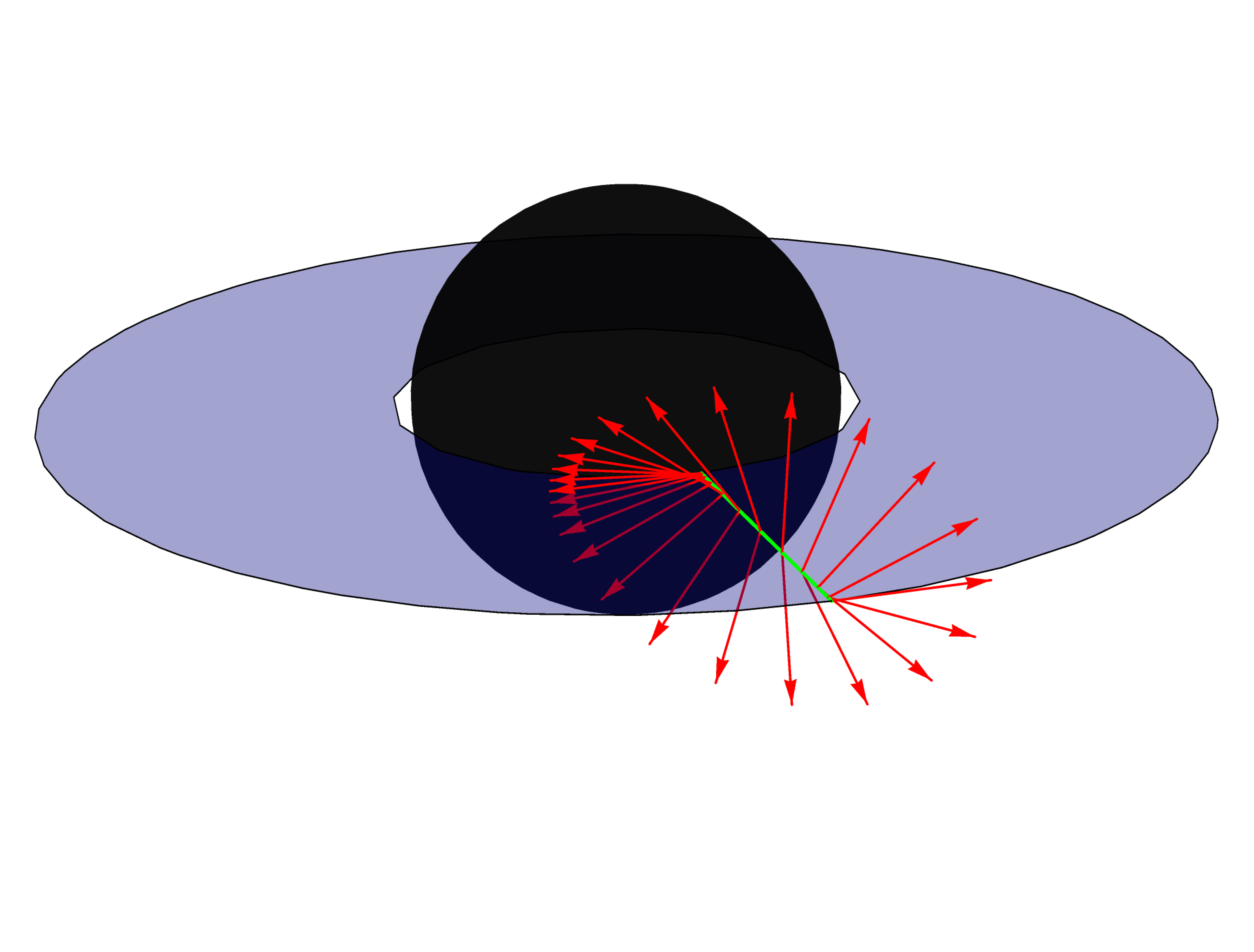}

		\caption{A set of null vectors from the family $(\tilde{\ell}_{ K})^a$ initially tangent to a photon surface in the photon region for different $K$. All vectors start from the equatorial plane, and follow the future directed null geodesic orbit which is identified as belonging to a photon surface. Blue disk: radial extent of the photon region in the equatorial plane. Green line: line of initial points for each of the null vectors. Red arrows: the null tangents, only the direction is important. As $K=0 \to \pi$, the initial tangent rotates from being anti-aligned to the rotation of the black hole, to being aligned with it, as the radius of the spherical orbit decreases. As $K=\pi \to 2\pi$, the vectors rotate back to anti-alignment, but on the other side of the equatorial disk. $M=1,a=0.9$. }
	\end{figure}

	We can use the limiting behaviour of the spin coefficients to identify an orbit which passes near the north pole $\theta=0$, taking the limit $\theta \to 0$ in $\cos(K)$ gives a phase angle of $K=\pi/2$. Using this phase and $\theta\to 0$, we utilize the $\tilde{\epsilon}+\bar{\tilde{\epsilon}}$ equation as an implicit equation for the radius of the orbit $r$ which ``passes" through this polar angle, though no actual orbit can pass exactly through the poles in these coordinates. Numerically solving the equation gives the following parameters to trace an orbit that passes close to the $\theta=0$ axis,
	
	\begin{equation}
		K=\frac{\pi}{2}, \ \theta=0, \ r = 2.88321774192...,
	\end{equation}
	where the value $K=\pi/2$ identifies the sphere which passes through the $z$ axis, but to find an actual set of initial conditions so that a null geodesic emitted from the equatorial plane passes near the $z$ axis, we find that the parameters required are:
	
	\begin{equation}
		K=\frac{86\pi}{60} \approx \sqrt{2}\pi, \ \theta=\pi/2, \ r = 2.88321774192...
	\end{equation}

	For each photon surface inside the photon region, the rulings of the surface are given by the null geodesics corresponding to a phase of $K$ and $-K$; the vectors initially tangent to these geodesics are $(\tilde{\ell}_K)^a$ and $(\tilde{\ell}_{-K})^a$, and have inner product:
	
	\begin{equation}
		(\tilde{\ell}_K)^{a} (\tilde{\ell}_{-K})_a=-4\sin^2K.
	\end{equation}
	Clearly the vectors are aligned to each other when $K=0,\pi~(n\pi ~\forall n\in \mathbb{Z})$, which are the circular photon surfaces in this context, otherwise they are not aligned. We can adapt the frame to the surface rulings by performing the above null rotation about $n$, and then a null rotation about $\tilde{\ell}^a$ with the same parameter $E=\exp(i K)$.

	\subsection{The Null Geodesic Equations for $\ell^a$}

Here we examine the geodesic equations for a null vector in the spacetime, and show that the conditions that identify the photon surfaces above in terms of spin coefficients can equivalently be explored via the geodesic equations directly.

For the null vector $\ell^a = dx^a/d\lambda$, tangent to a curve $x^a(\lambda)$ parameterized by the affine parameter $\lambda$, the geodesic equations can be written as the following 8 dimensional system (effectively first order equations since the second derivatives are the derivatives of the components of $\ell^a$):

\begin{equation}
	\frac{d\ell^a}{d\lambda} + \Gamma^a_{\ bc} \ell^b \ell^c=0, \ \frac{dx^a}{d\lambda} = \ell^a.
\end{equation}
The fully expanded initial data for the geodesic equations for the vector $\ell^a$ are

\begin{subequations}
	\begin{align}
		\frac{d\ell^1}{d\lambda}  &= -\frac{4Ma^2r\sin(K)\sin(2\theta)}{R^4\sqrt{Q}}, \\
		\frac{d\ell^2}{d\lambda}  &= \frac{1}{R^4}\left[Ma^2+a^2\cos^2(\theta)(M-r)+r(3a^2+2r(r-3M))\right.\\
		&~~~~\left.+a\sin(\theta)(4\cos(K)r\sqrt{Q}+a(r-M)\sin(\theta)\right], \\
		\frac{d\ell^3}{d\lambda}  &= \frac{1}{2R^4}\left[\cot(\theta)(a^2(5+\cos(2K)+(\cos(2K)-3)\cos(2\theta))\right. \\
		&~~~~~\left.+4r^2\cos^2(K)+8a\cos(K)\sin(\theta)\sqrt{Q}\right], \\
		\frac{d\ell^3}{d\lambda}  &= \frac{4\cot(\theta)\sin(K)}{R^4}\left(\cos(K)\csc(\theta)R^2+\frac{aR^2+r(r-2M))}{\sqrt{Q}}\right), \\ 
		\frac{dt}{d\lambda}  &= \sqrt{2}\left(\frac{a\cos(K)\sin(\theta)}{R}+\frac{a^2+r^2}{R\sqrt{Q}}  \right), \\ 
		\frac{dr}{d\lambda}  &= 0, \\ 
		\frac{d\theta}{d\lambda}  &= -\frac{\sqrt{2}\sin(K)}{R}, \\ 
		\frac{d\phi}{d\lambda}  &= \frac{\sqrt{2}}{R}\left(\cos(K)\csc(\theta)+\frac{a}{\sqrt{Q}}\right). \\ 
	\end{align}
\end{subequations}

Consider $\ell^a$ starting initially in the equatorial plane ($\theta=\pi/2$), with arbitrary null rotation angle $K$, we can simplify the equations above in terms of coordinate position $x^a(\lambda)$, with our null vector given by the null rotated $\ell^a$,

\begin{subequations}
	\begin{align}
		\left.\frac{d^2t}{d\lambda^2}\right|_{\theta=\pi/2} &= 0, \\
		\left.\frac{d^2r}{d\lambda^2}\right|_{\theta=\pi/2} &= \frac{2\Bigr((2a^2+r(r-3M) + 2a\cos(K)\sqrt{a^2+r(r-2M)}\Bigr)}{r^3}, \\
		\left.\frac{d^2\theta}{d\lambda^2}\right|_{\theta=\pi/2} &= 0, \\
		\left.\frac{d^2\phi}{d\lambda^2}\right|_{\theta=\pi/2} &= 0, \\ 
		\left.\frac{dt}{d\lambda}\right|_{\theta=\pi/2} &= \sqrt{2}\left(\frac{a\cos(K)}{r}+\frac{a^2+r^2}{r\sqrt{a^2+r(r-2M)}}  \right), \\ 
		\left.\frac{dr}{d\lambda}\right|_{\theta=\pi/2} &= 0, \\ 
		\left.\frac{d\theta}{d\lambda}\right|_{\theta=\pi/2} &= -\frac{\sqrt{2}\sin(K)}{r}, \\ 
		\left.\frac{d\phi}{d\lambda}\right|_{\theta=\pi/2} &= \sqrt{2}\left(\frac{\cos(K)}{r}+\frac{a}{r\sqrt{a^2+r(r-2M)}}\right). \\ 
	\end{align}
\end{subequations}
Along the null curve traced by the geodesic equations, $dr/d\lambda$ vanishes in the equatorial plane (actually everywhere in the spacetime). To guarantee that the curves stay on a surface of constant $r$, clearly we only need to choose initial data $(r,K(r))$ such that $d^2r/d\lambda^2|_{\theta=\pi/2}=0$, which corresponds exactly to the vanishing of $\tilde{\kappa}, \ \tilde{\epsilon}+\bar{\tilde{\epsilon}}$ there. Also observe that for the angles $K=0,\pi$, the equations for the coordinate $\theta(\lambda)$ are constant for all affine parameter values, where the null geodesic locations with $dr^2/d\lambda^2=0$ identify the two circular orbits in the equatorial plane.

\subsection{The photon region defined by $P(K,r,\theta)$}
Using our invariant null geodesic congruence $\tilde{\ell}^a$, we will identify the entire photon region parameterized by the null rotation parameter $K$, using the geodesic equations. From the previous analysis we have that the null rotated spin coefficients $\tilde{\kappa}$ and $\tilde{\epsilon}+\bar{\tilde{\epsilon}}$ take the following forms in terms of the invariant frame spin coefficients and null rotation parameter $K$:

	\begin{equation}
		\tilde{\kappa} = 2e^{iK}(2\epsilon + \rho + \cos(K)(2\beta + \tau)),
	\end{equation}

	\begin{equation}
		\tilde{\epsilon}+\bar{\tilde{\epsilon}} = 2\epsilon + \rho + 2\bar{\epsilon} + \bar{\rho} + e^{iK}(\beta + \bar{\beta}+ \tau)+ e^{-iK}(\beta + \bar{\beta}+ \bar{\tau}).
	\end{equation}
	Consider the combination of the real parts of $\tilde{\epsilon}$ and $\tilde{\kappa}$:
	
	\begin{equation}
		(\tilde{\epsilon}+\bar{\tilde{\epsilon}} )-  (\tilde{\kappa} +\bar{\tilde{\kappa}})  =2i\sin(K)\left[ 2(2\bar{\epsilon} +\bar{\rho})+ \cos(K)(4\bar{\beta}+\tau +\bar{\tau})+i\sin(K)(\tau-\bar{\tau})   \right],
	\end{equation}
	which tells us that the real parts of $\tilde{\epsilon}$ and $\tilde{\kappa}$ are equivalent for $K=0,\pi$. That is, affinely parameterized null geodesic behaviour is defined equivalently by the zero set of either spin coefficients real part for these choices of $K$. The vanishing of the real parts of $\tilde{\kappa}$ or $\tilde{\epsilon}$ for $K=0,\pi$ form the boundary of the photon region, as these choices correspond to the null vector being aligned to the outermost photon orbits or the innermost orbits (not just the circular ones). The expressions can be evaluated for $K=0,\pi$:
	
	\begin{equation}\label{eq:photon_0}
		\left.\tilde{\epsilon}+\bar{\tilde{\epsilon}}\right|_{K=0} = \frac{r \left(4 a  \sin (\theta ) \sqrt{r (r -2 M )+a ^2}+2 r (r -3 M )+3 a ^2\right)+a ^2 \cos (2 \theta ) (M -r )+M 
   a ^2}{\sqrt{2} R\sqrt{ Q^3}},
	\end{equation}

	\begin{equation}\label{eq:photon_pi}
		\left.\tilde{\epsilon}+\bar{\tilde{\epsilon}}\right|_{K=\pi} = \frac{r \left(-4 a  \sin (\theta ) \sqrt{r (r -2 M )+a ^2}+2 r (r -3 M )+3 a ^2\right)+a ^2 \cos (2 \theta ) (M -r )+M 
   a ^2}{\sqrt{2} R\sqrt{ Q^3}}.
	\end{equation}
	The general expression for any $K,$ fully expanded is 
	
	\begin{equation}
		\tilde{\epsilon}+\bar{\tilde{\epsilon}} =\frac{a R (a  \sin (2 \theta ) \sin (K)-4 r  \sin (\theta ) \cos (K))+r  \left(6 M  r -2 r ^2-3 a ^2\right)-2 M 
   a ^2 \cos ^2(\theta )+r  a ^2 \cos (2 \theta )}{\sqrt{2} R\sqrt{Q^3}}.
	\end{equation}
	
	Using the expressions for $K=0,\pi$ we can plot the boundary of the photon region as an implicit surface in the $(x,z)$ plane, shown in Figure \ref{fig:photoninterior}. For $K\in (0,\pi)$, every other axially symmetric slice of the photon region between the boundary curves is picked out. The complete invariant structure of the photon region is determined by the vanishing of the function $P$ defined in (\ref{eq:ps-bl}), writtenly in terms of invariantly defined functions:
	
	\begin{equation}\label{eq:psinvariant}
		P(K,r,\theta) \equiv 2\epsilon + 2\bar{\epsilon} + \rho + \bar{\rho} + \cos(K)\left(2\beta + 2\bar{\beta} + \tau + \bar{\tau}\right).
	\end{equation}
	Any frame defined invariantly by the procedure in the previous section will also detect the photon region in this way, and as this is a coordinate invariant scalar, we can rewrite the expression in any desired coordinate system. The above expression can be expanded in Boyer-Linquist coordinates:

	\begin{equation}\label{eq:psinvariant_expanded}
		P(K,r,\theta)=-\frac{2 \left(r  \left(4 a  \sin (\theta ) \cos (K) \sqrt{r  (r -2
   M )+a ^2}+2 r  (r -3 M )+3 a ^2\right)+a ^2
   \cos (2 \theta ) (M -r )+M  a ^2\right)}{\sqrt{r(r-2M)+a ^2} \left(2 r ^2+a ^2 \cos (2 \theta)+a ^2\right)^{3/2}}.
	\end{equation}

	As a comparison, consider the function derived in \cite{Grenzebach_photon_regions}, where the authors define a condition to identify the boundary of the photon region in Boyer Lindquist coordinates via the first integrals of the geodesic equations, leading to the condition
	
	\begin{equation}\label{eq:photoninequality}
		4 \left(a ^2 \cos ^2(\theta) (M -r )+r  \left(r  (r -3 M )+2
   a ^2\right)\right)^2 \le 16a^2r^2\sin^2(\theta)(a^2-2Mr+r^2),
	\end{equation}
	where the equality is on the boundary of the region. To show that our invariant identifies the boundary, in the same manner as the above inequality, we observe that the vanishing of the numerator of $P(K,r,\theta)$, splitting the expression apart, inserting $K=0, \pi$ (expressions only have a single sign difference, see equations (\ref{eq:photon_0}), (\ref{eq:photon_pi})) and squaring both sides, results in
	
	\begin{equation}\label{eq:proofepsilonzeros}
		(4ra\sin(\theta)\sqrt{a^2+r(r-2M)})^2 = \Bigr(2Ma^2\cos^2(\theta)-ra^2\cos(2\theta)-r(6Mr-2r^2-3a^2)\Bigr)^2,
	\end{equation}
	Substitute the left hand side of (\ref{eq:proofepsilonzeros}) into the right hand side of (\ref{eq:photoninequality}), then yields

	\begin{equation}\label{eq:subst}
		4 \left(a ^2 \cos ^2(\theta) (M -r )+r  \left(r  (r -3 M )+2
   a ^2\right)\right)^2 = \Bigr(2Ma^2\cos^2(\theta)-ra^2\cos(2\theta)-r(6Mr-2r^2-3a^2)\Bigr)^2,
	\end{equation}
	which after a quick expansion and simplification, the expressions (\ref{eq:photoninequality}) and (\ref{eq:subst})  are seen to be identical. Thus the zeros of our invariant at $K=0,\pi$ identifies the photon region boundary, in the same manner as the expressions in \cite{Grenzebach_photon_regions}, but also gives the entire region in terms of a parameterization in a single function, without the inequalities.


\subsection{Interior of the Photon Region}
\begin{figure}[!htb]
	\centering
	\includegraphics[scale=0.6]{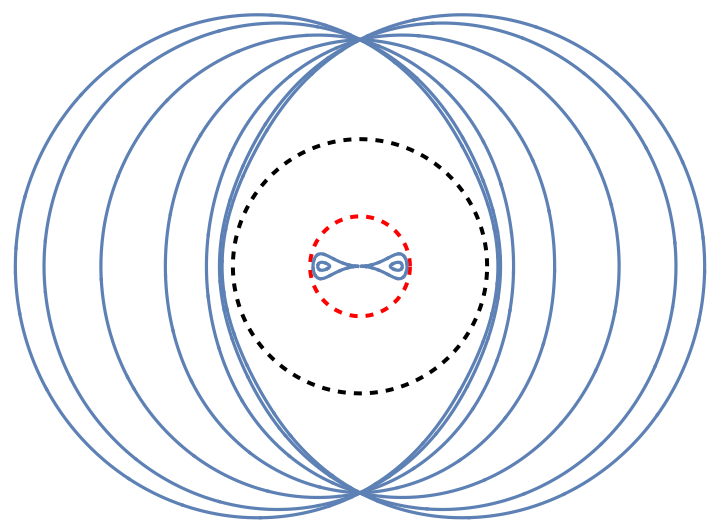}
	\caption{Interior foliation of the photon region by $P(K,r,\theta)=0$ for $K\in [0,\pi]$, $a=0.9$. Solid lines are the slices of the photon region in the interior and exterior of the horizon region. Black and red circles are the event horizon and Cauchy horizon respectively.}
	\label{fig:photoninterior}
\end{figure}

Not only does our function $P(K,r,\theta)$ define the boundary of the photon region, it also defines every two dimensional hypersurface in between the boundaries (see Figure \ref{fig:photoninterior} for a foliation of the photon region by $P$). In the interior of the Cauchy horizon $(r<r_{-})$, the frame is well defined and the photon surface invariant also identifies the boundary of the photon region in these interior coordinates. The center of the coordinates in the Boyer-Lindquist system are not particularly useful for analyzing the interior directly, as they are hiding the ring singularity at $(r=0,\theta=\pi/2$), but for any null geodesics which pass through the ring, from angles outside of the equatorial plane, the behaviour is well defined and tracked via the geodesic equations without issue. The only geodesics which can touch the ring singularity are the ones which are contained entirely in the equatorial plane in the inner region \cite{oneill2014}.

\begin{figure}[!htb]
	\centering
	\begin{minipage}{0.45\textwidth}
		\includegraphics[scale=0.6]{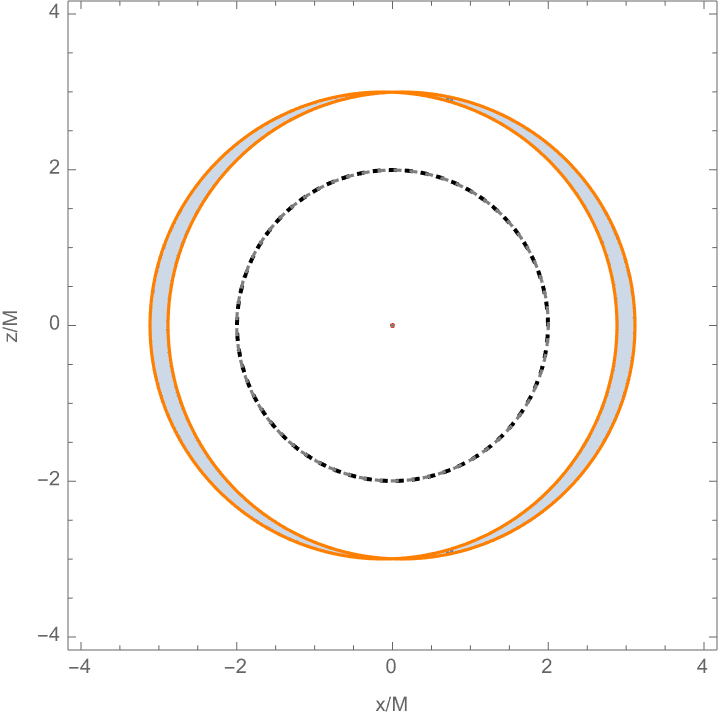}
	\caption{$a=1/10$.}
	\label{fig:photonregion_clean4}
	\end{minipage}
	\hfill
	\begin{minipage}{0.45\textwidth}
	\includegraphics[scale=0.6]{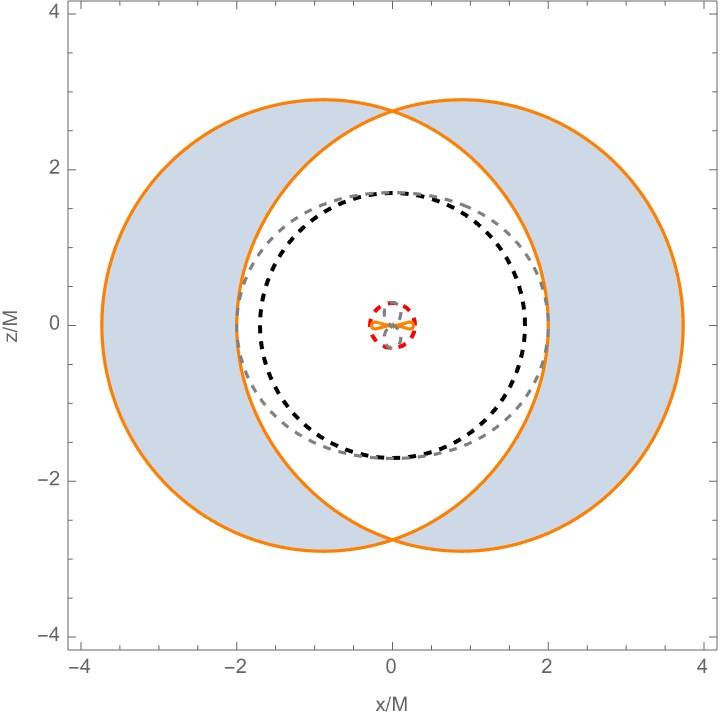}
	\caption{$a=1/\sqrt{2}$.}
	\label{fig:photonregion_clean3}
	\end{minipage}
	\caption{Photon region (blue), boundary of the photon region (orange), ergosurfaces (gray dashed), event horizon (black dashed), Cauchy horizon (red dashed). }
\end{figure}

\begin{figure}[!htb]
	\centering
	\begin{minipage}{0.45\textwidth}
		\includegraphics[scale=0.6]{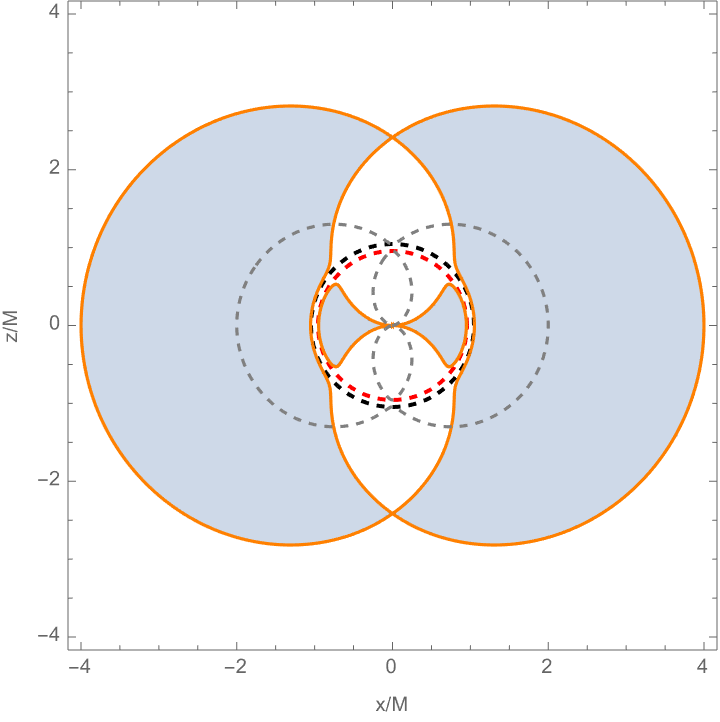}
	\caption{$a=0.999$.}
	\label{fig:photonregion_clean}
	\end{minipage}
	\hfill
	\begin{minipage}{0.45\textwidth}
	\includegraphics[scale=0.6]{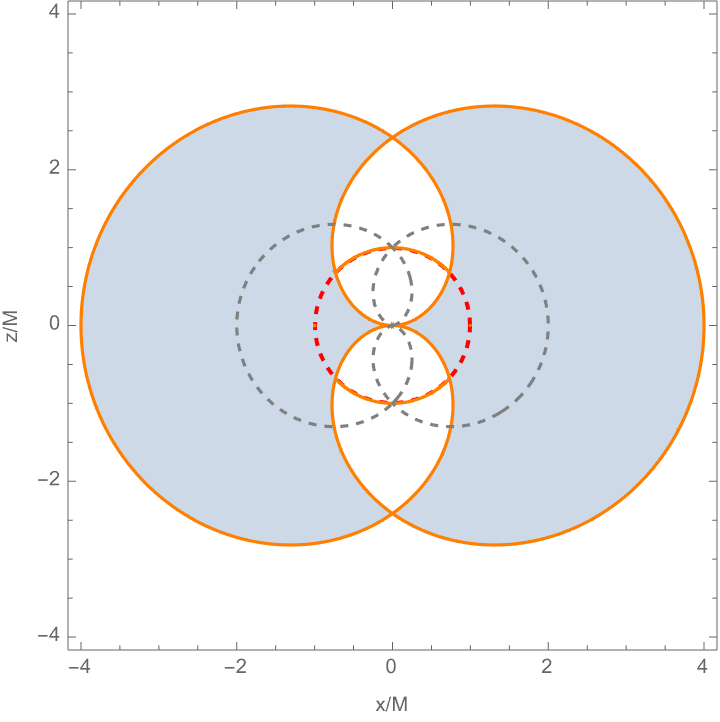}
	\caption{$a=1$.}
	\label{fig:photonregion_clean2}
	\end{minipage}
	\caption{Photon region (blue), boundary of the photon region (orange), ergosurfaces (gray dashed), event horizon (black dashed), Cauchy horizon (red dashed). }
\end{figure}

%
%
%
%
%
%
%
%

\section{First Integral Equations and Constants of Motion}\label{ch:firstintegralequations}

The separability of the Hamilton-Jacobi equation for the Kerr spacetime implies that there exists a conserved quantity of the action, which turns out to be Carter's constant, defined below. The Hamilton-Jacobi equations for the action $S$ of a null geodesic is given by

\begin{equation}
	g^{ab}\nabla_a S \nabla_b S=\frac{dS}{d\lambda}.
\end{equation}
In the NP formalism, replacing the metric with the tetrad and simplifying gives us the equivalent equation

\begin{equation}
	(\delta S)(\bar{\delta} S)=(DS )(\Delta S) + \frac12\frac{dS}{d\lambda}.
\end{equation}
Due to the Lorentz invariance of the metric, this expression is also frame independent. For null geodesics, the expression $dS/d\lambda=0$.

There exist two Killing symmetries of the Kerr metric. In addition, there is a Killing tensor, and together they define three of the constants of motion for any solution to the geodesic equations. The Killing vectors and tensor are given by,

\begin{subequations}
	\begin{align}
		\xi_t^a &= \{1,0,0,0 \}, \\
		\xi_\phi^a &= \{ 0,0,0,1\},\\
		\sigma_{ab} &= -a^2\cos^2(\theta) g_{ab} + R^2\left( e_{(2) a} e_{(2) b} + e_{(3) a} e_{(3) b}\right),
	\end{align}
\end{subequations}
where the basis vectors $e_{(2)a}$ and $e_{(3)a}$ are defined by
\begin{equation}
	e_{(2)a}=\frac{m_a+\bar{m}_a}{2}, \ e_{(3)a}=\frac{m_a-\bar{m}_a}{2i}.
\end{equation}

There is a fourth constant of motion, which is the inner product of the null vector, and is preserved by the metric trivially. It can be considered the Lagrangian of the particle with velocity $\ell^a$, for timelike particles. For the work here, we only are concerned with null trajectories, where the Lagrangian is always vanishing, and as a test of the stability of a numerical solution can be used as an error tracking function. The four constants of motion are the Lagrangian $\mathcal{L}$, the energy $E$, the angular momentum $L$, and the Carter constant $K_C$, and are defined by

\begin{subequations}
	\begin{align}
		&\mathcal{L} = g_{ab}\ell^a\ell^b,\\
	    &E = -\ell_a \xi_t^a, \\
	    &L = \ell_a \xi_\phi^a, \\
	    &K_C = \sigma_{ab} \ell^a\ell^b,
	\end{align}
\end{subequations}
and $\ell^a$ is a null geodesic vector. For the null geodesics specifically, we require $\mathcal{L}=0$ everywhere along the curve. The other constants can either be specified, or fixed by the initial data we generate by utilizing our invariantly defined $\ell^a$, or its null rotated form $\tilde{\ell}^a$ and the zeros of $P(K,r,\theta)$. 

If we generate initial data for the null curves as in the previous section, supposing that the null curves are emitted from the equatorial plane, parameterized by $K$, these constants of motion are initially fixed to be (where $r=r_0$ is a zero of $P(K,r,\pi/2)$)

\begin{subequations}
	\begin{align}
		&\mathcal{L}=0, \\
		&E= \frac{\sqrt{2}}{r_0}\left(a\cos(K)+\sqrt{a^2+r_0(r_0-2M)}\right), \\
		&L = \frac{\sqrt{2}}{r_0}\left( (a^2+r_0^2)\cos(K) + a\sqrt{a^2+r_0(r_0-2M)}\right), \\
		&K_C = 2r_0^2.
	\end{align}
\end{subequations}
The initial radius is computed from (in general) a sixth order polynomial in $r$, so we cannot find explicit expressions for the constants of motion. The values of the constants of motion are then fixed by initial data. We now produce a simple method for integrating the geodesic equations numerically and adapting the initial data to any photon surface.

\subsection{First Order Geodesic Equations}
Following the procedure by \cite{Chandra}, one can utilize the constants of motion to reduce the null geodesic equations to a set of nonlinear first order equations. Defining the parameter $\mathcal{Q}$:
\begin{equation}
	\mathcal{Q} \equiv K_C-(L-Ea)^2.
\end{equation}
We also redefine the affine parameter $\lambda$ by the relationship $d\lambda=R^2 ds $. The first order set of equations is then
\begin{subequations}\label{eq:geofirst}
	\begin{align}
		\dot{t}^2 &= \frac{1}{Q}\left( E\Bigr((r^2+a^2)^2-a^2 Q \sin^2(\theta)\Bigr) -2MaLr\right),\\
		\dot{r}^2 &= F(r),\\
		\dot{\theta}^2 &= \Theta(\theta),\\
		\dot{\phi}^2 &= 2MaEr+\frac{L}{\sin^2(\theta)}(r^2+a^2\cos^2(\theta)),
	\end{align}
\end{subequations}
where dot denotes differentiation with respect to the parameter $s$ along the null geodesic. The functions $F(r)$ and $\Theta(\theta)$ are 

\begin{subequations}
	\begin{align}
		F(r) &= Er^4+(a^2E^2-L^2-\mathcal{Q})r^2+2Mr(\mathcal{Q}+(L-aE)^2)-a^2\mathcal{Q}, \\
		\Theta(\theta) &= \mathcal{Q} + \left(a^2E^2-\frac{L^2}{\sin^2(\theta)}\right)\cos^2(\theta),
	\end{align}
\end{subequations}
 The $\dot{\theta}$ equation allows us to determine the maximum extent of the geodesics emitted from the equatorial plane, in terms of the constants of motion, which can further be related to the null rotation parameter $K$. Solving $\dot{\theta}=0$ gives the following result in terms of the constants of motion:

\begin{equation}
	\cos^2(\theta_{\rm min/max}) = \frac{(2Ea(Ea-L)-K_C) + \sqrt{K_C(K_C+4ELa)}}{2E^2a^2},
\end{equation}
and the $\pm$ roots of the cosine term determine the minimum and maxium angle $\theta$ of the orbit. 

\begin{figure}[htbp]
\centering

\begin{minipage}{0.32\textwidth}
  \centering
  \includegraphics[scale=0.03]{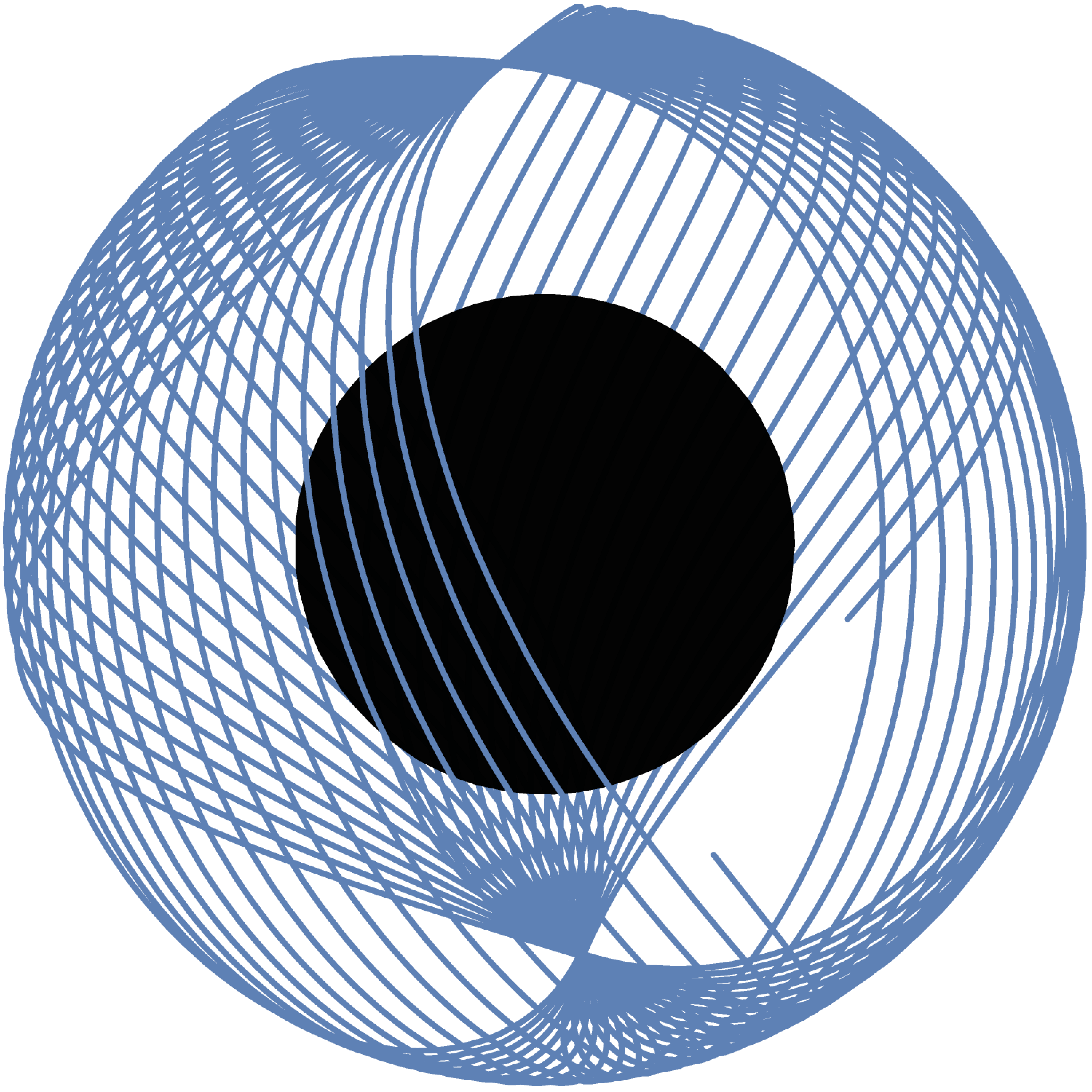}
\end{minipage}
\hfill
\begin{minipage}{0.32\textwidth}
  \centering
 	\includegraphics[scale=0.3]{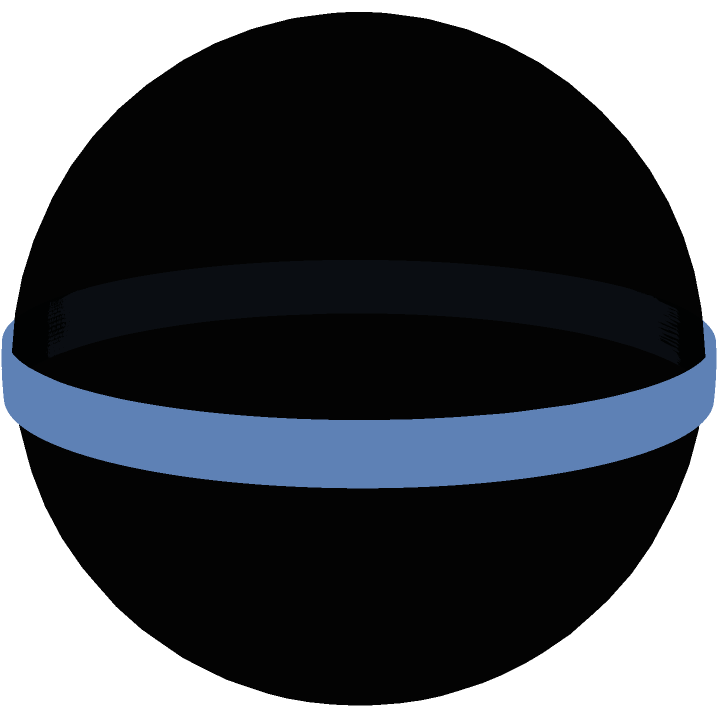}
	\label{fig:orbit1}
	\end{minipage}
\hfill
\begin{minipage}{0.32\textwidth}
  \centering
  \includegraphics[scale=0.3]{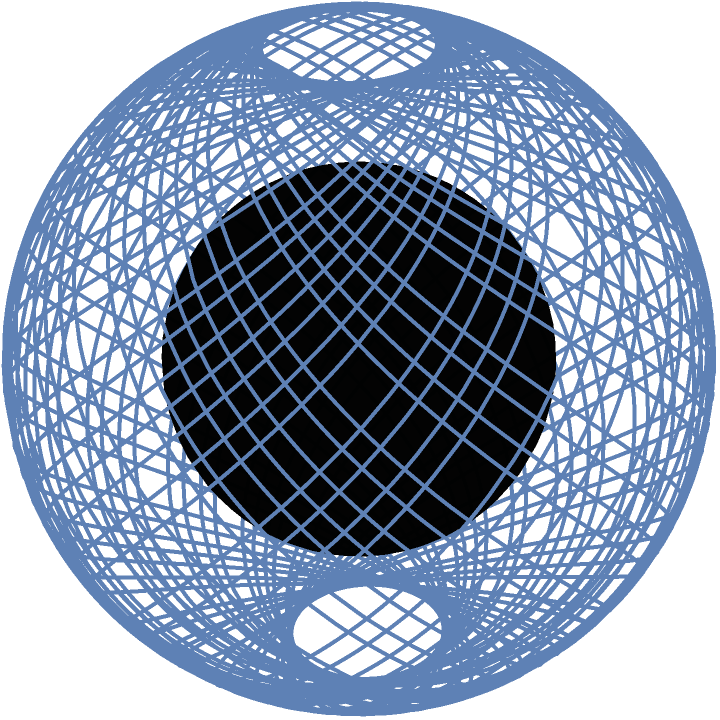}
	\label{fig:orbit3}
\end{minipage}

\vspace{0.5em} 

\begin{minipage}{0.32\textwidth}
  \centering
  \includegraphics[scale=0.03]{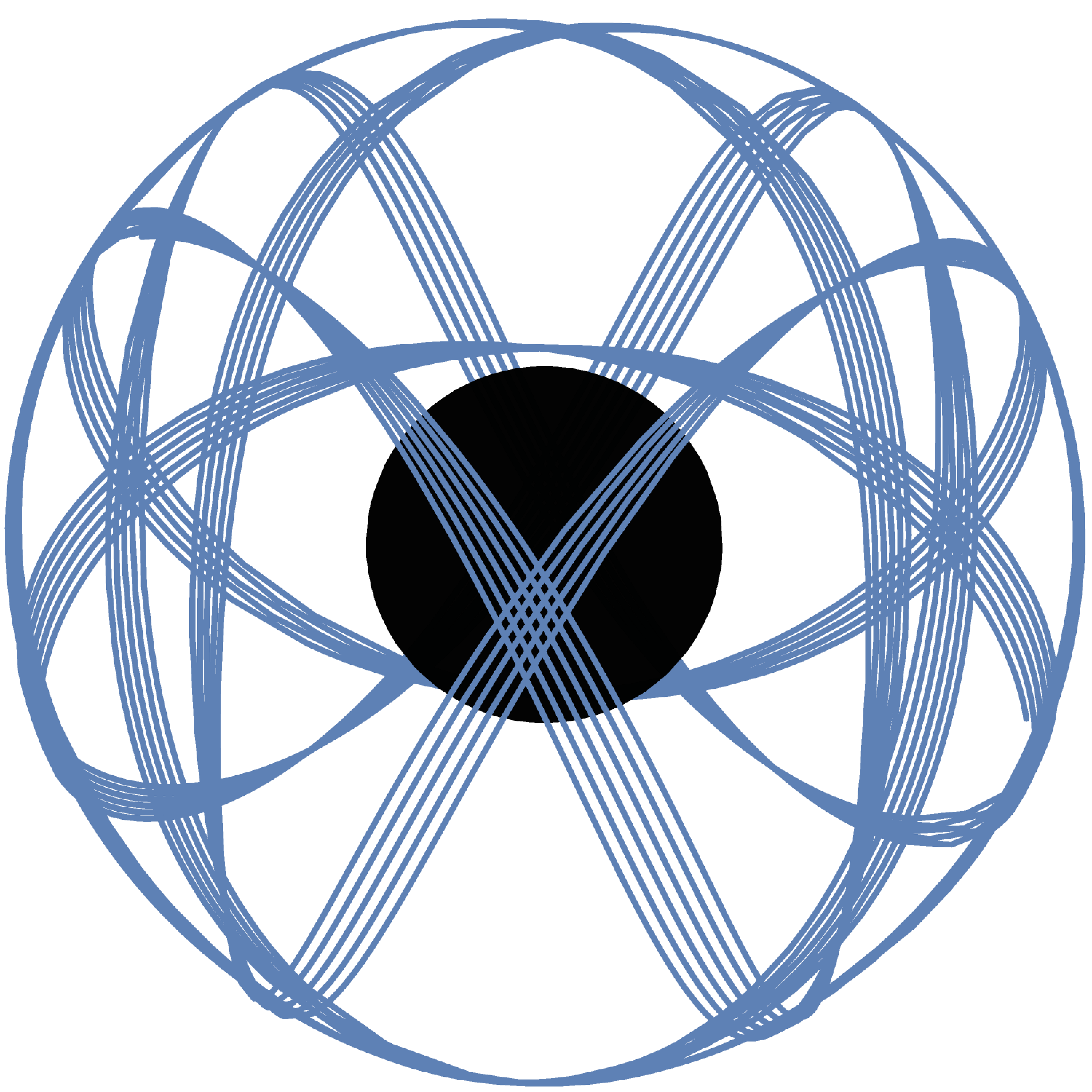}
	\label{fig:orbit4}
\end{minipage}
\hfill
\begin{minipage}{0.32\textwidth}
  \centering
  \includegraphics[scale=0.4]{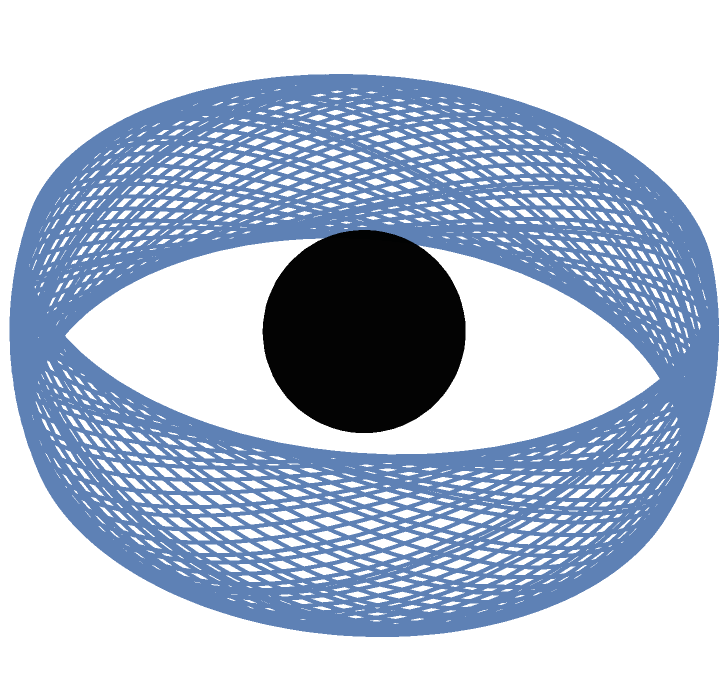}
	\label{fig:orbit5}
\end{minipage}
\hfill
\begin{minipage}{0.32\textwidth}
  \centering
  \includegraphics[scale=0.4]{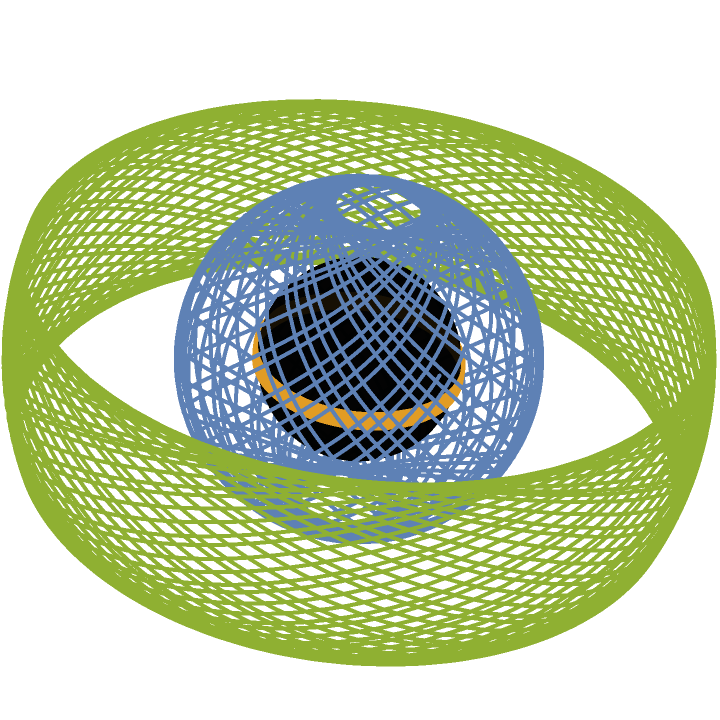}
	\label{fig:orbit6}
\end{minipage}

\caption{Various photon surfaces traced by geodesics. Initial data for $(r,\theta)$ were determined by $P(K,r,\theta)=0$ for different $K$. Each of the curves is a space filling curve, which traces the spherical hypersurfaces inside the photon region. Bottom right figure is three of the traced surfaces plotted together for scale comparison.}
\label{fig:ex24}

\end{figure}

Although exact solutions to these equations do exist, and can be written in terms of Weierstrass elliptic functions \cite{Hackmann2010}, running a numerical scheme to solve these equations is illustrative, and as a generalization of the method used here to identify photon surfaces in an arbitrary spacetime. Indeed, most sets of geodesic equations have no such analytic solutions (e.g., Szekeres spacetimes), so we need to be able to solve the geodesic equations numerically. One major issue one finds when trying to integrate the equations (\ref{eq:geofirst}) is that the $\dot{r}$ and $\dot{\theta}$ equations are quadratic in $\dot{r}$ and $\dot{\theta}$, and clearly when orbits reach a maximum or minimum of $r$ or $\theta$, a numerical solver won't be able to continue the evolution past these points since the sign of the square root of $F(r),\Theta(\theta)$ are needed. To solve this issue in the least obstructive way, we can simply differentiate the first order equations for $\dot{r}$ and $\dot{\theta}$, and obtain second order equations and add them to the numerical system:

\begin{subequations}
	\begin{align}
		\ddot{r}&=\frac12\frac{dR}{dr}, \\
		\ddot{\theta} &= \frac{1}{2}\frac{d\Theta}{d\theta}.
	\end{align}
\end{subequations}
When an extremum of $r$ or $\theta$ is reached, the second order term will allow the evolution to continue so that the square roots of the $\dot{r},\dot{\theta}$ equations remain real. We also define new variables to rewrite the geodesic equations as a first order system including $\dot{r} \equiv v, \ \dot{\theta}\equiv u$. The full system to integrate is now a six dimensional first order set of ordinary differential equations:

\begin{subequations}
	\begin{align}
		\dot{u} &= -K_C(r-M)-2Er(La-E(a^2+r^2)), \\
		\dot{v} &= L^2\cot(\theta)\csc^2(\theta)-E^2a^2\cos(\theta)\sin(\theta), \\
		\dot{t} &= \frac{1}{Q}\left( E\Bigr((r^2+a^2)^2-a^2 Q \sin^2(\theta)\Bigr) -2MaLr\right), \\
		\dot{r} &= u, \\
		\dot{\theta} &= v, \\
		\dot{\phi} &= 2MaEr+\frac{L}{\sin^2(\theta)}(r^2+a^2\cos^2(\theta)).
	\end{align}
\end{subequations}
To choose initial data which is guaranteed to be on a photon surface somewhere in the photon region,  we utilize the null vector which is aligned to the photon surfaces, $\tilde{\ell}^a$. For an affinely parameterized initial data set, we have that

	\begin{equation}
		\left.\frac{dx^a}{d\lambda}\right|_{\lambda=0}  = \left.\tilde{\ell}^a\right|_{s=0},
	\end{equation}
	so that reparameterized to the parameter $s$, we get the initial data:
	\begin{equation}
		\left.\frac{dx^a}{ds}\right|_{s=0} = \left.R^2  \tilde{\ell}^a\right|_{s=0}.
	\end{equation}
	Since the first order equations above only require initial data for $\ddot{r}=\dot{u}$ and $\ddot{\theta}=\dot{v}$, in the coordinate frame, the initial data is given directly by the components of $\tilde{\ell}^a$:
	
	\begin{equation}
		\dot{r}(0)=R^2\ell^2 = 0,
	\end{equation}
	where $\ell^2$ is the $r$ component of $\ell^a$, and for $\dot{\theta}$, the initial data is given by
	\begin{equation}
		\dot{\theta}(0) = R^2\ell^3 = 
		R^2\frac{\sqrt{2}\sin(K)}{r(0)^2+\cos^2(\theta(0))},
	\end{equation}
	where $\theta(0)$ and $r(0)$ are the roots of $P(K,r,\theta)$ for some value of $K$, and $\ell^3$ is the $\theta$ component of $\ell^a$. Choosing the initial data to lie in the equatorial plane, the angle $K$ is then the alignment of the initial data with respect to the equatorial plane, and this picks out the direction in which the null geodesic is emitted. Figure \ref{fig:ex24} shows some numerical plots of geodesics emitted from the equatorial plane tangent to a photon surface.

%
%
%

\section{Discussion}

We have presented a systematic, invariant approach to characterizing the Kerr spacetime, which has been discussed before, and has been expanded in greater detail here, with a focus on new approach to the photon region. The applications of the invariant characterization to the photon region in the Kerr spacetime gives a new alternative approach, which is different than the traditional coordinate dependent approach of using the metric to determine effective potentials to identify the photon surfaces. We provide an invariantly defined function $P(K,r,\theta)$, in equations (\ref{eq:psinvariant}), (\ref{eq:psinvariant_expanded}) which identify all photon orbits, and is valid for any coordinate transformation of $P$. The invariant approach gives the family of photon orbits which comprise the entire photon region as a single parametric function defined by the local geometry. This approach does show that it is possible to use invariant techniques to discuss the photon surface structure, in addition to the previous explored work on describing the horizons and ergoregion invariantly. The photon surface definition still appears to work directly as a computational tool in at least the Kerr spacetime (and by extension, this also works in the same manner for Kerr-Newman-NUT spacetimes, but the analysis is not reported here). The expectation is that the definition of a photon surface and this approach applies to axially symmetric spacetimes in general, static or dynamic, which is the subject of future work. In other works, the quasinormal modes of the perturbed Kerr photon orbits were shown to be related to the process of superradiant scattering and gravitational wave scattering around the Kerr black hole, and our approach provides a tool for precisely defining the photon surfaces and null frames tangent to each of the surfaces, to facilitate analysis of those particular phenomena, and as an emphasis for the usefulness of invariant techniques in general.

\newpage
\appendix
\section{Some Transformations of the NP Scalars}
We list the relationships for the NP scalars under Lorentz transformations, utilizing the invariant frame relationships as the initial frame. A tilde denotes the transformed scalar, and non-tilde refers to the function in the invariant frame. In the invariant null frame, we use the relationships among the spin coefficients and NP scalars to write all quantities in terms of $\Psi_2,\rho,\beta,\epsilon,\tau$. Under a null rotation with $B=e^{iK}$, $K\in \mathbb{R}$, the scalars transform relative to the invariant frame to

\begin{minipage}{0.48\textwidth}
\begin{subequations}
	\begin{align}
		\tilde{\kappa} &= 2e^{iK}(2\epsilon+\rho + \cos(K)(2\beta + \tau)), \\
		\tilde{\rho} &= 2\epsilon + \rho + 2e^{iK}\beta + e^{-iK}\tau, \\
		\tilde{\sigma} &= e^{iK}(2\beta + \tau + e^{iK}(2\epsilon + \rho)), \\
		\tilde{\tau} &= \tau + 2e^{iK}\epsilon, \\
		\tilde{\nu} &= 0, \\
		\tilde{\mu} &= \rho, \\
		\tilde{\lambda} &=0, \\
		\tilde{\pi} &= \tau + e^{-iK}\rho, \\
		\tilde{\epsilon} &= 2\epsilon +\rho + \cos(K)(2\beta + \tau) + i\sin(K) \tau, \\
		\tilde{\beta} &= \beta + e^{iK}(\epsilon + \rho), \\
		\tilde{\gamma} &= \epsilon, \\
		\tilde{\alpha} &= \beta + e^{-iK}\epsilon.
	\end{align}
\end{subequations}
\end{minipage}
\hfill
\begin{minipage}{0.48\textwidth}
\begin{subequations}
	\begin{align}
		\tilde{\Psi}_0 &= 6B^2\Psi_2, \\
		\tilde{\Psi}_1 &= 3B \Psi_2, \\
		\tilde{\Psi}_2 &= \Psi_2, \\
		\tilde{\Psi}_3 &= 0, \\
		\tilde{\Psi}_4 &= 0.
	\end{align}
\end{subequations}
\end{minipage}

\vspace{0.5cm}
\noindent
Under a spin and a null rotation, with parameters $\Theta$ and $B=e^{iK}$ as above, respectively, the scalars transform to:

\begin{minipage}{0.48\textwidth}
\begin{subequations}
	\begin{align}
		\tilde{\kappa} &= 2e^{i(K+\Theta)}(2\epsilon+\rho + \cos(K)(2\beta + \tau)), \\
		\tilde{\rho} &= 2\epsilon + \rho + 2e^{iK}\beta + e^{-iK}\tau, \\
		\tilde{\sigma} &= e^{i(K+2\Theta)}(2\beta + \tau + e^{iK}(2\epsilon + \rho)), \\
		\tilde{\tau} &= e^{i\Theta}(\tau + 2e^{iK}\epsilon), \\
		\tilde{\nu} &= 0, \\
		\tilde{\mu} &= \rho, \\
		\tilde{\lambda} &=0, \\
		\tilde{\pi} &= e^{-i(K+\Theta)}(e^{iK}\tau + \rho), \\
		\tilde{\epsilon} &= \frac{i}{2}\left( D\Theta + e^{-iK}\delta \Theta +e^{iK}\bar{\delta} \Theta  +\Delta \Theta \right)\nonumber\\
		&~~~~ + 2\epsilon +\rho + \cos(K)(2\beta + \tau) + i\sin(K) \tau, \\
		\tilde{\beta} &= e^{i\Theta}\left(\frac{i}{2}\delta\Theta + \frac{i}{2}\Delta \Theta + \beta + e^{iK}(\epsilon + \rho)\right), \\
		\tilde{\gamma} &= \epsilon + \frac{i}{2}\Delta \Theta, \\
		\tilde{\alpha} &= \frac{1}{2}e^{-i(K+\Theta)}\left( i\Delta \Theta + ie^{iK}\bar{\delta}\Theta + 2e^{iK}\beta + 2\epsilon\right).
	\end{align}
\end{subequations}
\end{minipage}
\hfill
\begin{minipage}{0.48\textwidth}
\begin{subequations}
	\begin{align}
		\tilde{\Psi}_0 &= 6e^{2i(\Theta+K)}\Psi_2, \\
		\tilde{\Psi}_1 &= 3e^{i(\Theta+K)} \Psi_2, \\
		\tilde{\Psi}_2 &= \Psi_2, \\
		\tilde{\Psi}_3 &= 0, \\
		\tilde{\Psi}_4 &= 0.
	\end{align}
\end{subequations}
\end{minipage}

\section{Coordinate Transformations}
For the given plots of this work, the Boyer-Lindquist coordinates are considered as ``spherical" coordinates in the standard sense, and are related to the Cartesian coordinates in the standard way:

\begin{subequations}
	\begin{align}
		x &= r\cos(\phi)\sin(\theta), \\
		y &= r\sin(\phi)\sin(\theta), \\
		z &= r\cos(\theta).
	\end{align}
\end{subequations}

%
%
%
%
%
%
%

%
%
	\printbibliography	
	
\end{document}